# THE KMOS³ᴰ SURVEY: DEMOGRAPHICS AND PROPERTIES OF GALACTIC OUTFLOWS AT $z = 0.6 – 2.7$[*]

N. M. Förster Schreiber[1], H. Übler[1], R. L. Davies[1], R. Genzel[1,2], E. Wisnioski[1,3,4], S. Belli[1], T. Shimizu[1], D. Lutz[1], M. Fossati[1,5], R. Herrera-Camus[1], J. T. Mendel[1,3,4], L. J. Tacconi[1], D. Wilman[1,5], A. Beifiori[1,5], G. B. Brammer[6], A. Burkert[1,5], C. M. Carollo[7], R. I. Davies[1], F. Eisenhauer[1], M. Fabricius[1], S. J. Lilly[7], I. Momcheva[8], T. Naab[9], E. J. Nelson[1,10], S. H. Price[1], A. Renzini[11], R. Saglia[1,5], A. Sternberg[12], P. van Dokkum[13], & S. Wuyts[14]

[1]*Max-Planck-Institut für extraterrestrische Physik (MPE), Giessenbachstr.1, D-85748 Garching, Germany*
[2]*Departments of Physics & Astronomy, University of California, 94720 Berkeley, USA*
[3]*Research School of Astronomy & Astrophysics, Australian National University, Canberra, ACT 2611, Australia*
[4]*ARC Centre for Excellence in All-Sky Astrophysics in 3D (ASTRO 3D), Australia*
[5]*Universitäts-Sternwarte, Ludwig-Maximilians-Universität, Scheinerstr. 1, München, D-81679, Germany*
[6]*Cosmic Dawn Center, Neils Bohr Institute, University of Copenhagen, Juliane Maries Vej 30, DK-2100 Copenhagen, Denmark*
[7]*Department of Physics, Institute for Astronomy, ETH Zurich, CH-8093 Zurich, Switzerland*
[8]*Space Telescope Science Institute, Baltimore, MD 21218, USA*
[9]*Max-Planck-Institut für Astrophysik, Karl-Schwarzschildstr. 1, D-85748 Garching, Germany*
[10]*Harvard-Smithsonian Center for Astrophysics, 60 Garden Street, Cambridge, MA 02138, USA*
[11]*INAF-Osservatorio Astronomico di Padova, Vicolo dell'Osservatorio 5, I-35122 Padova, Italy*
[12]*School of Physics and Astronomy, Tel Aviv University, Tel Aviv 69978, Israel*
[13]*Astronomy Department, Yale University, New Haven, CT 06511, USA*
[14]*Department of Physics, University of Bath, Claverton Down, Bath, BA2 7AY, United Kingdom*

## ABSTRACT

We present a census of ionized gas outflows in 599 normal galaxies at redshift 0.6<$z$<2.7, mostly based on integral field spectroscopy of H$\alpha$, [NII], and [SII] line emission. The sample fairly homogeneously covers the main sequence of star-forming galaxies with masses 9.0<log($M_*$/M$_\odot$)<11.7, and probes into the regimes of quiescent galaxies and starburst outliers. About 1/3 exhibits the high-velocity component indicative of outflows, roughly equally split into winds driven by star formation (SF) and active galactic nuclei (AGN). The incidence of SF-driven winds correlates mainly with star formation properties. These outflows have typical velocities of ~450 km s$^{-1}$, local electron densities of $n_{\rm e}$~380 cm$^{-3}$, modest mass loading factors of ~0.1−0.2 at all galaxy masses, and energetics compatible with momentum driving by young stellar populations. The SF-driven winds may escape from log($M_*$/M$_\odot$)≲10.3 galaxies but substantial mass, momentum, and energy in hotter and colder outflow phases seem required to account for low galaxy formation efficiencies in the low-mass regime. Faster AGN-driven outflows (~1000−2000 km s$^{-1}$) are commonly detected above log($M_*$/M$_\odot$)~10.7, in up to ~75% of log($M_*$/M$_\odot$)≳11.2 galaxies. The incidence, strength, and velocity of AGN-driven winds strongly correlates with stellar mass and central concentration. Their outflowing ionized gas appears denser ($n_{\rm e}$~1000 cm$^{-3}$), and possibly compressed and shock-excited. These winds have comparable mass loading factors as the SF-driven winds but carry ~10 (~50) times more momentum (energy). The results confirm our previous findings of high duty cycle, energy-driven outflows powered by AGN above the Schechter mass, which may contribute to star formation quenching.

---

# 1. INTRODUCTION

## 1.1. Low Galaxy Formation Efficiency

Galaxies have been fairly inefficient in forming stars from the available baryons over cosmic time. Studies linking the distributions of galaxies and dark matter halos indicate that the galactic stellar fraction is below ~20% of the cosmic baryon abundance (e.g., Madau et al. 1996; Baldry et al. 2008; Conroy & Wechsler 2009; Guo et al. 2010; Moster et al. 2010, 2013; Yang et al. 2012; Behroozi et al. 2013a,b). The maximum value is reached at a halo mass of about $\log(M_h/M_\odot)\sim 12$ (or galaxy stellar mass $\log(M_*/M_\odot)\sim 10.5$), and drops to <10% on either side of this mass. Galactic winds driven by supernovae and massive stars have long been proposed to explain the low baryon content of low mass halos (e.g., Dekel & Silk 1986; Efstathiou 2000). The decreasing efficiency of galaxy formation at the high mass tail may be caused by less efficient cooling and accretion of baryons in massive halos (e.g., Rees & Ostriker 1977; Keres et al. 2005; Dekel & Birnboim 2006; Feldmann et al. 2016). Alternatively, or additionally, efficient outflows driven by the energetic output of accreting massive black holes in active galactic nuclei (AGN) may quench star formation at and above the Schechter stellar mass $\log(M_S/M_\odot)\sim 10.8$ (e.g., Peng et al. 2010b; Behroozi et al. 2013b) either by ejecting gas already in the galaxy, or by preventing gas from coming into it (e.g., di Matteo et al. 2005; Croton et al. 2006; Bower et al. 2006, 2015; Hopkins et al. 2006; Cattaneo et al. 2006; Ciotti & Ostriker 2007; Somerville et al. 2008; Fabian 2012; Pillepich et al. 2018).

## 1.2. Feedback from Star Formation

Feedback in the form of outflows has long been observed in nearby starburst galaxies, characterized by exceptionally intense star formation activity $\gtrsim 0.1$ $M_\odot$ yr$^{-1}$ kpc$^{-2}$ (e.g., Heckman 2002; Veilleux et al. 2005; Heckman & Thompson 2017 for reviews). As the best-studied case, M82 exemplifies the richness of the outflow phenomenon and in particular its multiphase nature, with its wind being traced in hot $\gtrsim 10^6$ K X-ray emitting gas, warm ~$10^4$ K gas in UV to far-infrared (far-IR) line emission, and colder atomic and molecular gas as well as dust in the near-/mid-IR, mm, and radio observations (e.g., Heckman et al. 1990; Shopbell & Bland-Hawthorn 1998; Lehnert et al. 1999; Hoopes et al. 2005; Veilleux et al. 2009; Contursi et al. 2013; Beirão et al. 2015; Leroy et al. 2015). Outflow velocities in local starburst winds typically are ~ 500 km s$^{-1}$ and mass outflow rates ($\dot{M}_{\rm out}$) are roughly comparable to the star formation rates (SFRs), albeit with a wide range depending on the galaxy and the outflow phase (e.g., Martin 2005; Veilleux et al. 2005; Chisholm et al. 2015; Heckman et al. 2015). In addition to the complications due to their multi-phase nature, the mass outflow rates and energetics of the winds are affected by considerable uncertainties related notably to the geometry, physical conditions, and filling or covering factor of the outflowing material.

At intermediate and high redshift out to z ~ 3, galactic winds in (non-AGN) star-forming galaxies (SFGs) and even post-starburst galaxies have been primarily observed through rest-UV to optical interstellar absorption lines and nebular emission lines (e.g., Franx et al. 1997; Pettini et al. 2000; Shapley et al. 2003; Tremonti et al. 2007; Weiner et al. 2009; Steidel et al. 2010; Rubin et al. 2010a,b; Genzel et al. 2011; Bordoloi et al. 2011, 2014; Bouché et al. 2012; Diamond-Stanic et al. 2012; Kacprzak et al. 2012; Kornei et al. 2012; Martin et al. 2012; Newman et al. 2012a,b; Talia et al. 2012, 2017). In contrast to the local universe, outflows are ubiquitous among higher redshift SFGs, not surprisingly given the rapid evolution in galactic SFR $\propto$ $(1+z)^\alpha$ with $\alpha \sim 2.5 - 3$ (e.g., Madau & Dickinson 2014).

For momentum- or energy-driven outflows powered by star formation, theoretical considerations and numerical simulations predict a power-law dependence of the mass loading factor $\eta = \dot{M}_{\rm out}$/SFR on galaxy stellar mass and circular velocity, with slopes in the range $-1/3$ to $-2/3$, and $-1$ to $-2$, respectively (e.g., Chevalier & Clegg 1985; Kauffmann et al. 1993; Benson et al. 2003; Murray et al. 2005; Oppenheimer & Davé 2008; Dutton 2012; Muratov et al. 2015). Observations of local star formation-driven winds are broadly consistent with these scalings, implying more efficient outflows in lower-mass galaxies (e.g., Heckman et al. 2015; Chisholm et al. 2017), but very few quantitative constraints exist at higher redshift.

## 1.3. Feedback from AGN

In the local Universe, preventive AGN feedback has been directly observed in the so-called radio mode in central cluster galaxies driving jets into the intra-cluster medium, which in turn create rarified, buoyant bubbles in the circumgalactic medium (CGM; see reviews by Heckman & Best 2014; McNamara & Nulsen 2007; Fabian 2012). AGN feedback also acts by expelling gas from the nuclear regions, traced as ionized gas winds in Seyfert 2 galaxies (e.g., Cecil et al. 1990; Veilleux et al. 2005; Westmoquette et al. 2012; Rupke & Veilleux 2013; Harrison et al. 2014), and powerful neutral and ionized gas outflows from obscured AGNs in late-stage, gas-rich mergers and Type 1 quasars (e.g., Fischer et al. 2010; Feruglio et al. 2010; Sturm et al. 2011; Rupke & Veilleux 2013; Veilleux et al. 2013; Arribas et al. 2014; Rupke et al. 2017).

At high redshift, ejective AGN feedback in the "QSO mode" has also been observed in rest-UV absorption lines or rest-optical nebular emission lines in broad absorption line quasars (e.g., Arav et al. 2001, 2008, 2013; Korista et al. 2008), in Type 2 AGN (e.g., Alexander et al. 2010; Nesvadba et al. 2011; Cano Díaz et al. 2012; Harrison et al. 2012b, 2016; Förster Schreiber et al. 2014; Genzel et al. 2014; Brusa et al. 2015a,b, 2016; Zakamska et al. 2016; Wylezalek & Zakamska 2016; Talia et al. 2017), and in radio galaxies (e.g., Nesvadba et al. 2008). At all redshifts, AGN-driven outflows typically have high velocities of ∼ 1000 – 2000 km s$^{-1}$, and inferred mass loading factors systematically higher than those of star formation-driven winds (e.g., Fiore et al. 2017), albeit again with substantial uncertainties. Most studies at high redshift focussed on galaxies hosting luminous QSOs or X-ray selected samples.

In terms of energetics, QSO mode feedback is in principle capable of driving the gas content of a gas-rich, high-redshift galaxy into the CGM or even intergalactic medium (IGM; e.g., Murray et al. 2005; Heckman 2010; Fabian 2012; Hopkins et al. 2016). However, luminous AGNs near the Eddington limit are rare. QSOs constitute <1% of the star-forming galaxy population in the same mass range (e.g; Boyle et al. 2000), have short lifetimes ($t_{QSO}$ ∼ $10^7 - 10^8$ yr, much less than the Hubble time $t_H$; Martini 2004), and thus have low duty cycles compared to galactic star formation processes ($t_{SF}$ ∼ $10^9$ yr; Hickox et al. 2014). From the broader perspective of galaxy evolution, it is then unclear how the rare and highly variable QSO mode of ejective feedback can quench star formation in massive galaxies, at least on the long run.

Perhaps consistent with this concern, the recent observational literature is inconclusive whether the QSO mode does (e.g., Cano Díaz et al. 2012; Alatalo et al. 2015; Brusa et al. 2015b; Tombesi et al. 2015; Carniani et al. 2016; Cheung et al. 2016; Wylezalek & Zakamska 2016) or does not (e.g., Harrison et al. 2012a; Mullaney et al. 2012; Santini et al. 2012; Balmaverde et al. 2016; Bernhard et al. 2016; Bongiorno et al. 2016) have much effect in regulating galaxy growth and star formation shutdown. Based on simulations, Pillepich et al. (2018) and Nelson et al. (2018) have proposed that low-Eddington ratio, kinetic feedback from accreting massive black holes may be more efficient in quenching star formation through a preventive feedback in the CGM (see also Bower et al. 2017).

### 1.4. Exploring Feedback from the Perspective of the Normal Galaxy Population

Our studies of star formation and AGN feedback at $z$∼1−3 focus on outflows among the normal galaxy population as a whole. These studies were part of the SINS/zC-SINF and KMOS$^{3D}$ surveys carried out with the SINFONI and KMOS near-IR integral field unit (IFU) spectrometers at the Very Large Telescope (VLT). Our samples were selected primarily based on redshift, stellar mass, near- or mid-IR magnitudes, and, for SINS/zC-SINF, also on SFR from large multi-wavelength surveys with deep near-IR or optical source extractions and a high mass completeness. Such a selection provides a fair census of the underlying population of "main sequence" (MS) SFGs (Förster Schreiber et al. 2009, 2018; Mancini et al. 2011, Wisnioski et al. 2015). With these IFU samples, we investigated the incidence and properties of ionized gas outflows traced by broad Hα, [NII]λλ6548,6584, and [SII] λλ6716,6731 line emission as a function of galaxy parameters. This approach is different from pre-selecting samples that probe subsets of the full galaxy population potentially undergoing short evolutionary phases (e.g., luminous QSOs), and is better suited to address the nature and role of feedback in a population- and time-averaged sense.

In Genzel et al. (2011) and Newman et al. (2012a,b, hereafter N12a, N12b), we showed from SINS/zC-SINF adaptive optics (AO) assisted data that spatially extended, broad (velocity width of FWHM ∼ 400 − 600 km s$^{-1}$) Hα+[NII]+[SII] emission commonly arises from massive star-forming clumps and from galactic disks above a SFR surface density threshold of $\Sigma_{SFR}$ ∼ 0.5−1 M$_\odot$ yr$^{-1}$ kpc$^{-2}$ (see also Davies et al. 2019). This component plausibly traces the launching sites of galactic winds driven by stellar feedback, traced on larger scales by interstellar absorption lines (e.g.; Shapley 2003; Steidel et al. 2010; Weiner et al. 2009; Rubin et al. 2010b; Bordoloi et al. 2014).

In Förster Schreiber et al. (2014, hereafter FS14), we reported the discovery of even broader (FWHM ∼ 1000 − 2500 km s$^{-1}$) ionized gas emission originating from the central few kpc in six of seven massive (log($M_*$/M$_\odot$) >10.9) $z$ ∼ 2 MS SFGs from the SINS/zC-SINF survey, mostly observed with AO. Together with further observations of high-mass SFGs with SINFONI+AO, this broad and centrally concentrated emission has now been spatially resolved in about 10 galaxies, indicating an intrinsic diameter of ∼ 2 – 3 kpc (FS14; Genzel et al. 2014; R. L. Davies et al., in preparation). The fact that this nuclear broad emission component is present in the forbidden [NII] and [SII] lines, and is extended on kpc-scales, excludes that the broad emission comes from a virialized, parsec-scale AGN broad-line region (BLR; e.g., Netzer 2013). The velocities, size, and spectral properties imply that the source of this broad emission component is not gravitationally bound and represents an outflow in the kpc-scale narrow-line region (NLR) associated with an (obscured) AGN (e.g., Cecil et al. 1990; Netzer 2013; Westmoquette et al. 2012; Rupke & Veilleux 2013).

The broad emission in all these galaxies is characterized by elevated ratios of broad to narrow Hα fluxes ~ 0.3−1 (Genzel et al. 2011, 2014; N12a; N12b; FS14). Depending on the local electron density of the outflowing gas emitting the broad component, $n_{e,br}$, these ratios could imply substantial mass outflow rates and mass loading factors of the SF- and AGN-driven winds. Assuming a simple spherical or biconical outflow with a constant velocity $v_{out}$, a radius $R_{out}$, and a constant electron density (Genzel et al. 2011), the mass loading factor can be expressed as:

$$\eta \sim 0.9 \times [F_{br}/F_{na}(H\alpha)] \times \left(\frac{n_{e,br}}{100 \text{ cm}^{-3}}\right)^{-1} \times \left(\frac{v_{out}}{400 \text{ km s}^{-1}}\right) \times \left(\frac{R_{out}}{3 \text{ kpc}}\right)^{-1}. \quad (1)$$

Noting the high fraction of nuclear AGN-driven outflows among the most massive MS SFGs in the SINS/zC-SINF sample, the next questions are on the incidence and parameter dependence of the associated broad emission component. In Genzel et al. (2014; hereafter G14), we expanded our study to 110 $z$~1−2.7 galaxies, half of which observed during the 1st year of the KMOS$^{3D}$ survey, with an emphasis on the high mass end. We found that the incidence of fast nuclear outflows increases rapidly with stellar mass, reaching 62±12% (1σ) at log($M_*/M_\odot$)>10.9, and including several sub-MS galaxies with SFR estimates and rest-frame *UVJ* colors consistent with their being quiescent (see also Belli et al. 2017). This incidence of nuclear outflows is almost twice that of AGN identified from X-ray, mid-infrared (mid-IR), or radio indicators among the sample studied, as well as in other surveys (e.g., Reddy et al. 2005; Papovich et al. 2006; Daddi et al. 2007; Brusa et al. 2009; Bongiorno et al. 2012; Hainline et al. 2012; Mancini et al. 2015).

In the present paper, we take advantage of the now much larger KMOS$^{3D}$ survey, with >700 galaxies observed in Hα+[NII], to revisit the demographics and physical properties of galactic outflows at $z$~0.6−2.7. With 525 well-detected sources in KMOS$^{3D}$, complemented with smaller sets from SINS/zC-SINF and other near-IR slit spectroscopic studies (Kriek et al. 2007; Newman et al. 2014; Wuyts et al. 2014; Barro et al. 2014; van Dokkum et al. 2015), the new sample analyzed here represents a substantial increase in size by a factor of 5.5 compared to G14, with a wider and more complete coverage of galaxy parameter space. It permits a more detailed investigation and leads to more robust conclusions about the incidence and trends in outflow properties among the overall galaxy population.

The paper is organized as follows. Section 2 describes the galaxy sample and the measurements used to characterize the presence and properties of outflows. Section 3 presents the results on the incidence and the separation between SF- and AGN-driven outflows adopted for the analysis. Section 4 discusses the trends in outflow incidence and physical properties as a function of galaxy parameters. The paper is summarized in Section 5. Throughout, we adopt a Chabrier (2003) stellar initial mass function and a ΛCDM cosmology with $H_0 = 70$ km s$^{-1}$ and $\Omega_m = 0.3$. Magnitudes are given in the AB photometric system.

## 2. GALAXY SAMPLES AND DATA SETS

### 2.1. Galaxy Sample Assembly

Our main objective is to determine the incidence and properties of outflows at $z$ ~1 − 3 as a function of galaxy parameters, identified through the Hα, [NII], and [SII] line emission. Spatially-resolved IFU data are very powerful for this, allowing removal of velocity broadening due to gravitational (orbital) motions across the galaxies, thus increasing the contrast between, and S/N of, the broad outflow emission and the narrow component from star formation. These considerations drove the choice of near-IR IFU and spectroscopic samples included in the present study, resulting in a total of 599 galaxies spanning 0.6 < $z$ < 2.7 (hereafter "full sample") from the work listed in this subsection. The selection criteria from all these samples were largely primarily based on stellar mass and redshift (with only a minority involving an additional cut related to star formation activity). This approach minimizes biases towards sub-population selection of AGNs or starbursts, which could probe the outflow phenomenon in more extreme, rarer short-lived phases.

### 2.1.1. KMOS$^{3D}$ IFU Sample

The vast majority of the galaxies (525 of 599, or 88%) are taken from the 5-year KMOS$^{3D}$ survey with the multi-IFU instrument KMOS at the VLT (Wisnioski et al. 2015, and in preparation). The survey strategy emphasized sensitive observations of individual sources and wide coverage of galaxy parameters in stellar mass, SFR, and colors in three redshift slices at $z$ ~ 0.9, $z$ ~ 1.5, and $z$ ~ 2.2 (with Hα observed in the *YJ*, *H*, and *K* bands, respectively).

The KMOS$^{3D}$ targets were drawn from the 3D-HST source catalog (Skelton et al. 2014; Momcheva et al. 2016), a near-IR grism survey with the *Hubble Space Telescope* (*HST*) in the CANDELS *HST* imaging survey fields (Grogin et al. 2011; Koekemoer et al. 2011) and with extensive X-ray to radio multi-wavelength data. The selection criteria for KMOS$^{3D}$ were a *K*-band magnitude $K_{AB} \leq 23$ mag, $\log(M_*/M_\odot) > 9.0$, and a secure and sufficiently accurate redshift (either from the $R\sim130$ grism spectra or from $R>300$ optical/near-IR slit spectra) to ensure sky line avoidance for the lines of interest[2]. These criteria reduced biases towards brighter, bluer, and more intensely star-forming objects.

Up to the end of 2017 December, a total of 725 galaxies were observed in KMOS$^{3D}$ for their H$\alpha$+[NII] emission[3]. These data have a median seeing-limited FWHM resolution of 0.5″, are split roughly equally between the *YJ*, *H*, and *K* bands, and have typically ~8h on-source integration times (with a median of 5, 8, and 9h in *YJ*, *H*, and *K*, respectively). The high H$\alpha$ detection fraction of ≈79% is nearly constant across all bands; it increases to ≈91% among SFGs (defined as having a SFR relative to that of the MS at the source's redshift and stellar mass $\Delta$MS = log(SFR/SFR$_{MS}$) $> -0.85$ dex) and is an appreciable ≈28% for quiescent galaxies with $\Delta$MS $< -0.85$ dex.

We culled the objects for the present analysis among the detected sources with signal-to-noise ratio per spectral channel at H$\alpha$ of S/N > 3, and excluded cases where strong residuals from telluric lines affect the interval encompassing a few 1000 km s$^{-1}$ around the H$\alpha$+[NII] complex (see Section 2.5.1 for more details). The latter criterion is particularly important given the large velocity extent of the outflow emission and for optimizing the quality of the spectra. This yielded the set of 525 galaxies considered here (219 in *YJ*, 131 in *H*, 175 in *K* band), with median integration time of 8.2h (ranging from 1.2 to 28.8h).

### 2.1.2. SINS/zC-SINF IFU Sample

We included 47 galaxies from the SINS/zC-SINF seeing-limited and AO-assisted surveys with the single-IFU instrument SINFONI at the VLT (Förster Schreiber et al. 2009, 2018; Mancini et al. 2011). The SINS/zC-SINF targets were drawn from a collection of parent $z\sim1.5$–2.5 samples selected on the basis of their *K*-band, 4.5 μm, or optical magnitudes and/or colors, with accurate optical spectroscopic redshifts, and an expected observed integrated H$\alpha$ flux $\geq 5 \times 10^{-17}$ erg s$^{-1}$ cm$^{-2}$ (or equivalently a SFR $\gtrsim 10$ M$_\odot$ yr$^{-1}$). Of the 84 objects targeted for H$\alpha$ in the initial seeing-limited observations (typical FWHM of 0.6″), 74 were detected, and a representative subset of 35 were followed-up at high resolution with AO (median FWHM of 0.17″) and with deep integrations (median integration time of 6.0h, ranging from 2 to 23h). As extensively discussed in the references above, the SINS/zC-SINF galaxies probe well the $z\sim2$ SFG population over two orders of magnitude in stellar mass and SFR, although with biases in the low-mass regime towards bluer colors and higher specific SFRs, stemming from the parent surveys and the H$\alpha$ flux detectability. We further included here two $z\sim2$ sources observed with SINFONI+AO that were not part of the original SINS/zC-SINF sample; one of them, J0901+1814, was in our previous G14 study (and will be discussed in more detail by R. L. Davies et al., in preparation).

For the study of outflows here, we kept the SINS/zC-SINF galaxies observed in AO and natural seeing that have S/N > 3 per spectral channel at H$\alpha$ and no strong sky line residuals around H$\alpha$+[NII], and that do not overlap with the KMOS$^{3D}$ sample. The inclusion of these sources adds high-quality IFU data, especially towards lower masses at $z \sim 2$ where the main KMOS$^{3D}$–based set has sparser sampling and resolves more poorly these smaller galaxies.

### 2.1.3. High-Mass Spectroscopic Samples

To further boost the numbers of high-mass but rarer galaxies, necessary for robust statistics in this regime, we also included objects from the following sets with seeing-limited near-IR spectroscopic data: i) five $\log(M_*/M_\odot) \gtrsim 10.7$, $z = 1.4 - 2.3$ galaxies, initially selected from their optical redshift, observed with the LUCI multi-object spectrograph (MOS) at the Large Binocular Telescope (LBT; G14; Newman et al. 2014; Wuyts et al. 2014), including EGS-13011166 observed in slit-mapping mode (Genzel et al. 2013); ii) six emission line galaxies from the *K*-selected sample at $\log(M_*/M_\odot) \gtrsim 10.7$ and $2 < z < 2.5$ of

---

[2] No preference was given to targets with line emission in the 3D-HST grism data, which also provided both continuum absorption- and emission line-based redshifts with typically $\sim 700 - 1000$ km s$^{-1}$ accuracy. The combination of deep near-IR (rest-optical) source detection/extraction, and rest-optical absorption- and emission-based redshifts from rest-optical features is a key factor in including galaxies that have a faint or low equivalent width H$\alpha$ and very red colors. Such a selection makes resulting samples less biased against high levels of dust extinction and/or dominant evolved stellar populations compared to photometric selections based on optical brightness and colors, and optical spectroscopic redshifts.

[3] The survey was completed in 2018 March, with a final number of 740 unique targeted galaxies (E. S. Wisnioski et al., in preparation).

Kriek et al. (2007, 2008), observed with the GNIRS spectrograph on the Gemini South telescope and with VLT/SINFONI (see also G14); iii) the 16 of 25 star-forming compact massive galaxies with $\log(M_*/M_\odot) \gtrsim 10.7$ and $2.0 < z < 2.5$ from the 3D-HST survey presented by van Dokkum et al. (2015) and Barro et al. (2014) observed with the multi-slit MOSFIRE or the single-slit NIRSPEC spectrographs on the Keck II telescope, and which are not part of the KMOS$^{3D}$, SINS/zC-SINF, and LUCI samples. Integration times for these data sets are typically between 1h and 4h. Further details of the selection and observations can be found in the references given above. These spectroscopic samples will hereafter be referred to as "LUCI," "K07," and "vD15/B14."

## 2.2. Reference Galaxy Population

To place the full sample in the broader context of the galaxy population, we constructed a reference sample from the 3D-HST source catalogs. We selected all objects in the same $0.6 < z < 2.7$ range, and to the same $K_{AB} \leq 23$ mag and $\log(M_*/M_\odot) > 9.0$ limits as employed for KMOS$^{3D}$. These limits also encompass the ranges and selection criteria of the other spectroscopic sets included in the outflow sample. Quality cuts were further applied to exclude objects with unreliable photometry or grism spectra (e.g., due to strong contamination by bright neighbours) but sources with a photometric redshift only were included.

## 2.3. Galaxy Stellar and Size Properties

The stellar masses and SFRs were derived from the optical to near-IR broad- and medium-band spectral energy distributions (SEDs) of the galaxies, supplemented with mid-/far-IR photometry from $Spitzer$/IRAC between 3 and 8 μm, $Spitzer$/MIPS at 24μm, and $Herschel$/PACS at 70, 100, and 160 μm whenever available (which is the case for the vast majority of the galaxies as they mostly lie in the CANDELS/3D-HST and COSMOS survey areas). The derivation followed procedures as described by Wuyts et al. (2011). In brief, the optical to mid-IR SEDs[4] were fitted with Bruzual & Charlot (2003) population synthesis models, adopting the Calzetti et al. (2000) reddening law, solar metallicity, and a range of star formation histories (including constant SFR and exponentially declining SFRs with varying $e$-folding timescales). The SFRs from these SED fits were adopted or, for objects observed and detected in at least one of the 24 to 160 μm $Spitzer$/MIPS or $Herschel$/PACS bands, from rest-UV + IR luminosities through the Herschel-calibrated ladder of SFR indicators of Wuyts et al. (2011). For consistency, we used these derivations for the reference sample and all objects from the full sample within the CANDELS/3D-HST fields (KMOS$^{3D}$, LUCI, vD15/B14). For the SINS/zC-SINF and K07 sample, we used the values obtained through similar modeling procedures given by Förster Schreiber et al. (2009, 2018), Mancini et al. (2011), and Kriek et al. (2007, adjusted to our adopted Chabrier IMF).

The use of the ladder of SFR indicators is important because of the wide range in SFRs spanned by our sample, reaching $\gtrsim 50$ M$_\odot$yr$^{-1}$. In this regime, the SED-based derivations (SFR$_{SED}$) systematically underestimate the total intrinsic SFRs as measured from the UV+IR luminosities (SFR$_{UV+IR}$) because of increasing amounts of extinction, and of saturation of the reddening of SEDs related to the dust and sources distribution (see, e.g., Santini et al. 2009; Wuyts et al. 2011; and references therein). We stress however that more than 90% of the full sample lie in fields with MIPS and PACS observations, and that ~90% of the objects at SFR > 50 M$_\odot$yr$^{-1}$ have a SFR$_{UV+IR}$ determination. As shown by Wuyts et al. (2011) for the recipes we are using here, the average SFR$_{UV+IR}$ versus SFR$_{SED}$ relationship is monotonic. We verified that the relative trends in outflow properties and the conclusions discussed in this paper are unaffected by using the SFR$_{SED}$ values for all galaxies, because their ranking in SFR is overall preserved.

The galaxy sizes were obtained from two-dimensional Sérsic (1968) profile fits to HST $H$-band imaging, based on the GALFIT code (Peng et al. 2010a). For all KMOS$^{3D}$, LUCI, and vD15/B14 galaxies, as well as for the reference sample, the effective radii $R_e$ were taken from van der Wel et al. (2012; see also Lang et al. 2014). The sizes for the K07 objects are those presented by Kriek et al. (2009). For all SINS/zC-SINF galaxies with HST near-IR imaging, we used the results of Tacchella et al. (2015); for the other objects (12), we adopted the SINFONI-based Hα sizes, which provide reasonable approximations of the rest-optical continuum sizes (see Nelson et al. 2012, 2016; Förster Schreiber et al. 2018). The $R_e$ values adopted here refer to the major axis half-light radii.

The samples span a fairly wide redshift range over which the $M_*$–SFR and $M_*$–$R_e$ relations evolve significantly. Assuming that galaxies mainly grow along these relations, with up and down excursions due to fluctuations in accretion rate, (mostly minor) mergers, and in-situ radial transport (such as in "compaction events"; e.g., Zolotov et al. 2015; Tacchella et al. 2016), it is useful to consider their location relative to the MS and $M_*$–$R_e$ relations at the same mass and redshift. We thus computed for each galaxy its offset from the MS and from the $M_*$–$R_e$. The MS offset, $\Delta MS = \log(SFR/SFR(M_*,z))$ was calculated using the MS parametrization of Whitaker et al. (2014). Similarly, we defined $\Delta MR_e = \log(R_e/R_e(M_*,z))$ adopting the mass-size relation for SFGs and the (small) color correction between observed $H$-band and rest-frame 5000Å sizes from van der Wel et al. (2014).

---

[4]The SEDs cover observed-frame $U$ to 8μm wavelengths, sampled with 15 to 43 broad- and medium-band photometry for the 566 galaxies (94%) that lie in the 3D-HST/CANDELS and the COSMOS fields, and $U/B$ to $K$ band sampled with 5 to 10 broad-band photometry for the other 33 objects; details on the filter sets and photometry can be found in the references given in Section 2.1.

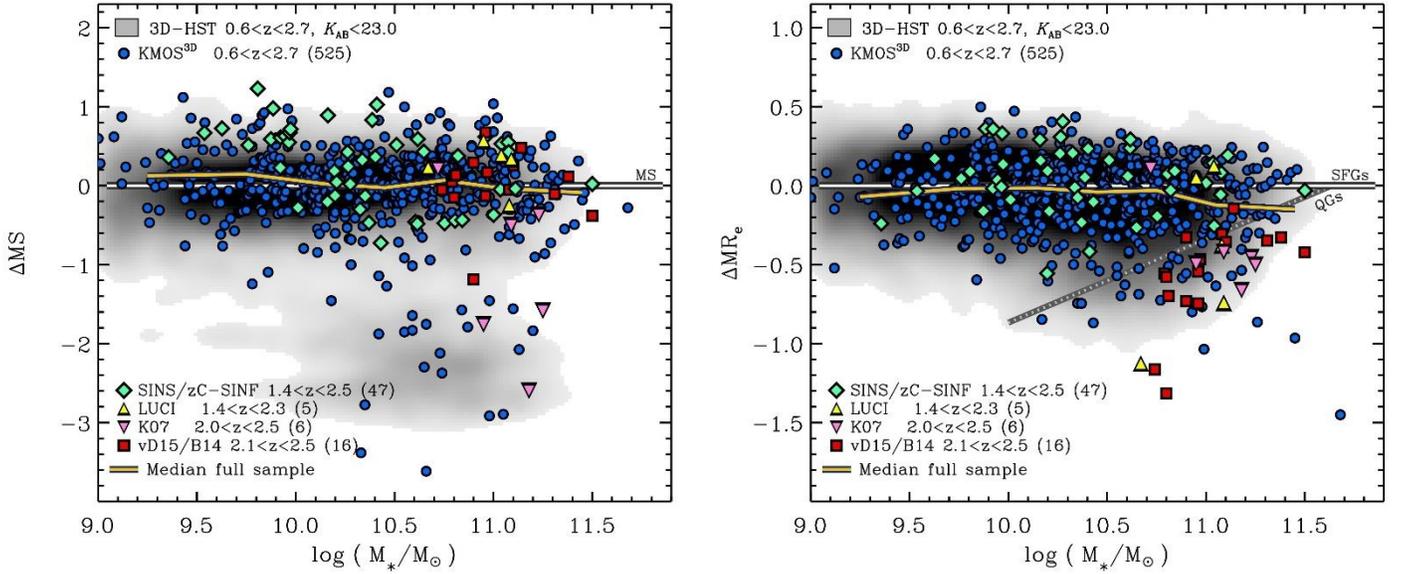

**Figure 1.** Distribution in stellar and size properties of the full sample of 599 galaxies assembled to study the demographics and properties of galactic outflows at $0.6 < z < 2.7$. The sample is compared to the underlying distribution of galaxies from the 3D-HST source catalog (Skelton et al. 2014; Momcheva et al. 2016) in the same redshift range, and with the same $\log(M_*/M_\odot) > 9.0$ and $K_{AB} < 23$ mag cuts as applied for the KMOS$^{3D}$ survey that dominates the full sample. Different symbols identify subsets taken from different parent IFU and slit spectroscopic samples, as described in Section 2.1 and labeled in the plots; the legend also gives the corresponding redshift ranges and numbers of galaxies from each subset. The density distribution of the reference 3D-HST population is shown in grey colors on a linear scale. *Left:* Stellar mass versus logarithmic offset in SFR from the MS at the redshift and mass of each galaxy, $\Delta MS = \log(SFR/SFR(M_*,z))$, using the parametrization of Whitaker et al. (2014); the black-white horizontal line shows $\Delta MS = 0$. *Right:* Stellar mass versus logarithmic offset in effective major axis radius at rest-frame 5000Å from the mass-size relation of SFGs at the redshift and mass of each galaxy, $\Delta MR_e = \log(R_e/R_e(M_*,z))$, using the fits of van der Wel et al. (2014). The black-white horizontal line shows the $\Delta MR_e = 0$, and the grey-dotted line indicates the mass-size relation for quiescent galaxies normalized to that of SFGs. In both plots, the black-yellow line indicates the median distribution of the sample galaxies along the vertical axis as a function of stellar mass. The full sample covers fairly homogeneously the SFG population in both $\Delta MS$ and $\Delta MR_e$ over the entire $9.0 \lesssim \log(M_*/M_\odot) \lesssim 11.5$ range, extends into the "starburst" and "quiescent" regimes above and below the MS, and also probes compact massive galaxies.

### 2.4. The Full Sample in Context

Figure 1 shows the distribution of the full sample of 599 galaxies in the $M_*$–$\Delta MS$ and $M_*$–$\Delta MR_e$ planes compared to that of the reference galaxy population from 3D-HST, with objects from different parent spectroscopic samples distinguished by different symbols. The sample spans wide ranges of $9.0 < \log(M_*/M_\odot) < 11.7$, $-3.62 < \Delta MS < 1.23$, and $-1.45 < \Delta MR_e < 0.50$ that cover well the underlying galaxy population. In particular, the coverage in $\Delta MS$ and $\Delta MR_e$ probes fairly homogeneously the bulk of SFGs within roughly $\pm 0.6$ dex and $\pm 0.3$ dex of the normalized MS and mass-size relations, respectively. Though more sparse, the sample also extends above the MS into the regime of "starburst outliers" ($\Delta MS > 0.6$ dex; e.g., Rodighiero et al. 2011), and below the MS down to the regime of quiescent galaxies. The overall median $\Delta MS$ is +0.04 dex, with only minor variations as a function of mass of $|\Delta MS| < 0.09$ dex at $\log(M_*/M_\odot) > 10$ and about +0.13 dex at lower masses (driven by the lower H$\alpha$ detection rate for low-mass $z \sim 2$ targets), well within the scatter of the MS of SFGs ($\sim 0.3$ dex; e.g., Rodighiero et al. 2011; Whitaker et al. 2014; Speagle et al. 2014). In size, the median $\Delta MR_e$ is $-0.04$ dex for the full sample, with similar or smaller offset at fixed mass for $\log(M_*/M_\odot) \lesssim 11$ and about $-0.13$ dex at higher masses (driven by the vD15/B14 subset that was selected by compactness), again all within the scatter in the mass-size relation for SFGs (~0.2 dex; van der Wel et al. 2014).

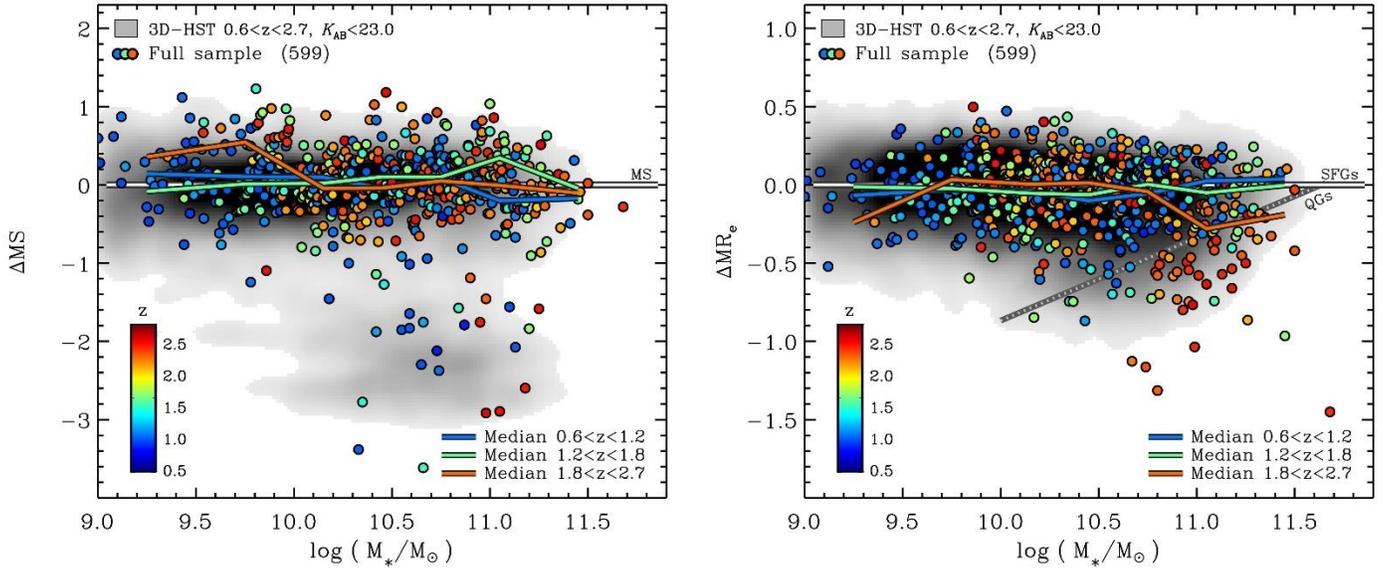

**Figure 2.** Distribution of the full sample and the reference galaxy population, in the same properties as in Figure 1 but now with circles representing the sample galaxies color-coded by redshift according to the color bar in the panels. *Left:* Stellar mass versus logarithmic offset in SFR from the MS. *Right:* Stellar mass versus logarithmic offset in effective major axis radius at rest-frame 5000Å from the mass-size relation of SFGs. In both plots, the colored lines indicate the median distributions of the sample galaxies along the vertical axis as a function of stellar mass for three redshift slices as labeled in the panels. The parameter space coverage and running median values are very similar for each redshift slice. The most important deviations are for the highest $z$ slice, with sparser and more biased sampling at lower mass, and larger proportion of compact high-mass galaxies, resulting from the selection criteria of the parent IFU and slit spectroscopic samples as explained in Section 2.4.

Figure 2 plots the sample in the same $M_*-\Delta$MS and $M_*-\Delta$MR$_e$ planes, now distinguishing galaxies by their redshift. Splitting the data in the three redshift intervals $0.6 < z < 1.2$, $1.2 < z < 1.8$, and $1.8 < z < 2.7$ (corresponding to H$\alpha$ observed in the $YJ$, $H$, and $K$ bands), the global median $|\Delta$MS$|$ and $|\Delta$MR$_e|$ are all less than 0.05 dex. The data cover well the same parameter ranges with no strong differentiation as a function of redshift except for the following trends. At $\log(M_*/M_\odot) \lesssim 10$, the sample is dominated by lower $z$ galaxies, which results primarily from the $K$-band magnitude cut of KMOS[3D], SINS/zC-SINF, and K07, and the mass selection criterion of the $z = 2 - 2.5$ vD15/B14 sample. Similar reasons together with H$\alpha$ detectability and the scarcity among the galaxy population explain the lack of low-mass objects well below the MS at any redshift. As noted above, the inclusion of the vD15/B14 galaxies leads to the higher proportion of compact massive galaxies at $z \gtrsim 2$. These trends are reflected in the more important deviations in the highest $z$ slice in the $\Delta$MS and $\Delta$MR$_e$ running median values. For intermediate masses, and for the two lowest $z$ slices across the full mass range, the running median values are fairly flat, with deviations of at most $\approx 0.35$ dex from the MS and $\approx 0.1$ dex from the mass-size relation of SFGs.

By design of the largely dominant KMOS[3D] survey, which aimed at a wide and uniform coverage with both a mass and $K$-band selection cut, and as a result of sensitivity of the data, the full sample naturally emphasizes more massive galaxies ($\log(M_*/M_\odot) \gtrsim 10.5$) compared to a purely mass-selected sample in the same $\log(M_*/M_\odot) > 9$ mass range. This is mostly apparent in the $z$ and $M_*$ distributions (from the selection criteria) and in SFR and rest-frame colors (where the H$\alpha$ detection rate drops most importantly among the objects with reddest colors and well below the MS; Belli et al. 2017). Nonetheless, the KMOS[3D] selection criteria and the high detection rate of ~90% of near-MS targets did not introduce preferential biases in a given redshift slice for the bulk of SFGs. The addition of the other, comparatively much smaller spectroscopic subsets hardly changes the resulting distributions. For the exploration of trends in incidence and properties of outflows as a function of galaxy parameters, the bulk of the full sample around the MS galaxies is little biased.

Compared to other concurrent surveys of rest-frame optical spectra of $z \sim 1-3$ galaxies based on multiplexed slit or IFU spectroscopy (e.g., KBSS, Steidel et al. 2014;

MOSDEF, Kriek et al. 2015; KROSS, Stott et al. 2016), the full sample here provides higher quality individual spectra (typical integration times in the other surveys were ~ 2h), and has an order of magnitude larger number of massive galaxies ($\log(M_*/M_\odot) > 10.8$) in the regime where star formation quenching by AGN feedback is expected to occur. Compared to the KASHz survey with KMOS of X-ray-selected AGN at $z \sim 0.6 - 1.7$ presented by Harrison et al. (2016), our sample primarily selected from stellar properties of the galaxies (stellar mass, and rest-frame optical luminosity through the $K$-band magnitude) enables us to assess the role and average properties of outflows among the galaxy population as a whole.

## 2.5. Data Analysis

The observations and data reduction are presented by Wisnioski et al. (2015; see also Davies et al. 2013) for the KMOS[3D] survey, Förster Schreiber et al. (2009, 2018) and Mancini et al. (2011) for the SINS/zC-SINF sample, Wuyts et al. (2014) and Newman et al. (2014) for the LUCI targets, Kriek et al. (2007) for the GNIRS and SINFONI data of the K07 sources, and by van Dokkum et al. (2015) and Barro et al. (2014) for the MOSFIRE and NIRSPEC data of the vD15/B14 sample. Our analysis used the fully reduced data sets, and we refer the reader to the references above for details about the observing strategies and reduction procedures. For the LUCI, K07, and vD15/B14 samples, we used the slit or combined slit+IFU integrated spectra as published. For the KMOS[3D] and SINS/zC-SINF galaxies, representing 95% of the data sets, we took advantage of the spatially-resolved information to extract spectra optimized for the identification and characterization of outflow emission, as described below. The reduced cubes have a spatial sampling of 0.2″ for KMOS, and 0.125″ and 0.05″ for SINFONI seeing-limited and AO data, respectively. The native spectral sampling is $\approx 40$ km s$^{-1}$ for KMOS and $\approx 35$ km s$^{-1}$ for SINFONI in both modes.

### 2.5.1. Spectral Extraction from Reduced IFU Data Cubes

We followed the methodology applied in our previous work on outflows (Shapiro et al. 2009; Genzel et al. 2011, 2014; Newman et al. 2012a,b; FS14). The fully reduced data cubes were first median-subtracted to remove continuum emission, which is well detected in most of the more massive SFGs of our sample. The data were then 4σ-clipped blueward and redward of the Hα+[NII] emission complex to remove telluric emission line shot noise. In some cases where a sky line was very close to the narrow (star formation-dominated) Hα emission, we interpolated over one, or up to at most four spectral channels. The cubes were then spatially smoothed with a Gaussian of FWHM between 3 to 4 pixels.

We fitted a single Gaussian line profile to the spectrum of each pixel to extract the smoothed velocity and velocity dispersion maps of the galaxy. A single Gaussian fit at the pixel level is mostly sensitive to the higher amplitude, narrower core of the line profile tracing emission from star-forming regions across galaxies, and, especially for the velocity of interest here, is little influenced by the broader and lower-amplitude outflow emission (see Förster Schreiber et al. 2018, Appendix C). The derived velocity field was then applied in reverse to the original data cube to remove large scale velocity gradients. This technique minimizes the impact of velocity broadening due to orbital motions in the final extracted spectra, and at the same time improves the S/N for detecting faint features and line wings. The method has obvious limitations for the compact sources with strong but unresolved inner velocity gradients; in such cases, the unresolved velocity gradients result in increased central velocity dispersions.

From the velocity-shifted cube of each galaxy, we extracted a spectrum covering the Hα, [NII], and [SII] lines (and also [OI]λ6300 for galaxies for which the line falls within the observed spectral band and is not contaminated by telluric lines). The spectra were typically integrated over the galaxies, or over the "nuclear" regions for the more extended galaxies; the typical extraction region was $\sim 0.5″-1.2″$ in diameter ($\sim 2-5$ kpc in radius). In the final selection of the galaxies, we rejected a few sources with very strong atmospheric contamination, such that only a small subset of the galaxies shows significant sky residuals in their Hα+[NII] profiles.

The final spectra for each galaxy were normalized to the peak amplitude at Hα and interpolated onto a common velocity sampling of 30 km s$^{-1}$. The quality of the spectra extracted from the data cubes naturally varies, owing to the variations in line flux and on-source integration times. With integrated Hα fluxes $\gtrsim 10^{-17}$ erg s$^{-1}$ cm$^{-2}$ (mean and median of $1.1\times 10^{-16}$ and $8.7\times 10^{-17}$ erg s$^{-1}$ cm$^{-2}$) and the long integration times (mean and median of 8h), the average and median S/N per spectral channel for Hα is ~15, with 56 galaxies having S/N > 30.

### 2.5.2. Spectral Stacking

Our aim of deriving the physical properties of outflows relies on multi-component fitting (broad + narrow) to the Hα line and [NII] and [SII] doublets as described in the next subsection. Although the overall S/N of the data is high, it is not sufficient to allow such fitting for all galaxies individually. We thus co-averaged spectra for different bins in global galaxy properties (with typically ~10 galaxies per bin, and up to ~60 depending on our purposes) following two approaches. In one approach, we computed the averaged spectrum with a uniform weighting for all galaxies in a bin. In a few cases, the resulting spectrum was of insufficient quality for reliable fitting of all features of interest because of the contribution of an individual lower S/N spectrum, which was then excluded. While providing a fair estimate of the average and being least affected by outliers, this choice obviously does not optimize the S/N of the stacked spectrum. Thus, in the second approach, we averaged the spectra weighting by S/N or (S/N)$^2$ to obtain the highest quality stacked spectra. We also compared results by further splitting up the objects in a bin. We found that these various methods make little difference, indicating that the properties of the stacked spectra are robust.

### 2.5.3. Spectral Fitting

Motivated by the earlier analyses of Ho et al. (1997), Genzel et al. (2011), and G14, we fitted multiple Gaussians to H$\alpha$, [NII]$\lambda\lambda$6548,6584, and whenever possible also [SII]$\lambda\lambda$6716,6731, with the following assumptions: (i) the systemic velocities and widths of the narrow component of all lines are identical, and likewise for the broad component, and (ii) the [NII]$\lambda$6548/$\lambda$6584 flux ratio is 0.326 (Storey & Zeippen 2000). The free parameters in the fitting were thus the FWHM of the narrow and the broad components (FWHM$_{na}$, FWHM$_{br}$), the velocity shift between the broad and narrow component centroids ($\Delta v_{br}$), the [NII]$\lambda$6584/H$\alpha$ flux ratio in the narrow and broad components ([NII]/H$\alpha_{na}$, [NII]/H$\alpha_{br}$), and the broad-to-narrow H$\alpha$ flux ratio $F_{br}/F_{na}$. In cases where we fitted the [SII] lines as well, the additional free parameters were the following flux ratios: [SII]$\lambda$6716+$\lambda$6731$_{na}$/H$\alpha_{na}$, [SII]$\lambda$6716+$\lambda$6731$_{br}$/H$\alpha_{br}$, [SII]$\lambda$6716/$\lambda$6731$_{na}$, and [SII]$\lambda$6716/$\lambda$6731$_{br}$. All narrow and broad components were always fit simultaneously, using the Python Markov Chain Monte Carlo sampler *emcee* (Foreman-Mackey et al. 2013). We explored the parameter space with on average 1000 walkers, and 2000 burn-in and run steps. Uncertainties were taken as the 68% percentile (1$\sigma$) bounds of the marginalized posterior distributions.

The S/N of the spectra does not always justify a 10 parameter fit, in which case we applied strong priors to a subset of the parameters, or fixed their values, based on results obtained from the higher S/N stacks. In particular, we generally restricted the FWHM$_{na}$ and FWHM$_{br}$ to values below and above 400 km s$^{-1}$; in some cases we fixed the values to FWHM$_{br}$ = 400 and 1000 km s$^{-1}$ for SF-driven and AGN-driven outflows, respectively. In all fits, we constrained the [SII] doublet ratio to the theoretically allowed range of 0.4315 < [SII]$\lambda$6716/$\lambda$6731 < 1.4484 (Sanders et al. 2016). Due to the small separation and weakness of the [SII] lines, the amplitudes of the narrow and broad components can be insufficiently constrained, especially for the cases of strong higher-velocity nuclear outflows. To break this degeneracy, we used a prior on [SII]$\lambda$6716/$\lambda$6731$_{na}$ of 1.34±0.03 obtained from a stack containing only sources *without* broad outflow emission. This ratio implies a local electron density within the HII regions of $n_{e,na}$ = 76$^{+24}_{-23}$ cm$^{-3}$, lower than recent estimates in the range 100–400 cm$^{-3}$ from multi-slit rest-optical spectroscopy of z ~ 1 – 2.5 SFGs (e.g., Masters et al. 2014; Steidel et al. 2014; Sanders et al. 2016; Kaasinen et al. 2017; Kashino et al. 2017). The difference may be due to contamination in single-component fits by broad emission from denser outflowing gas that is accounted for in our narrow+broad two-component fits (with $n_{e,br}$ ~ 380 cm$^{-3}$ for SF-driven outflows, and ~ 1000 cm$^{-3}$ for AGN-driven outflows; see Sections 4.1 and 4.2). When investigating the broad-to-narrow flux ratio distribution over the entire (binned) stellar mass versus $\Delta$MS plane, down to the weakest broad emission levels, we made the further simplifying assumption of a single broad line, comprising the sum of H$\alpha$ and [NII] (with the ratio denoted $F_{br}/F(H\alpha)_{na}$).

The assumption of a Gaussian line shape for the narrow component is justified in terms of the central limit theorem of many individual HII regions contributing to the integrated profile where large-scale velocity gradients have been removed, and the fitting results indicate it is adequate (see also, e.g., Genzel et al. 2011). It is less obvious or potentially wrong for the broad component, which in some cases appears to exhibit a blue/red asymmetry, such that the inferred line widths serve as a first order description.

In practice, given the S/N of the stacked data and as shown by G14, a broad component can be detected if its integrated flux is at least 10% that of the narrow component, and its width at least twice that of the narrow component. The average S/N is comparable across the stellar mass and $\Delta$MS ranges covered, such that the detectability of broad emission in terms of its relative flux fraction is roughly constant with these parameters. This assessment was verified quantitatively by adding broad Gaussian components of FWHM 400 and 1500 km s$^{-1}$ in H$\alpha$ and [NII] with varying amplitudes to stacked spectra in different mass bins (excluding those with strong detected broad components), and then analyzing the spectra as described above. In these stacks (of typically ~10 galaxies each), the minimum detectable broad component, in the sense of a significant and correct extraction of its width and flux (at the ≥ 3$\sigma$ level), is about 15-20% of the narrow component in terms of flux ratio, more or less flat across the mass range sampled by our data and similar for both widths (see G14, Figure 3). Broad components with fluxes down to about 10% that of the narrow component are still detectable but the inferred properties from spectral fitting are uncertain.

For individual galaxy spectra, the lowest detectable broad component obviously typically corresponds to higher broad flux levels, although the high S/N tail extends to similar values as the stacks such that similar limits as derived above are applicable in those cases. Defining a single reliable criterion for individual spectra is not straightforward because the detectability depends on the S/N as well as on the amplitude and width of the broad component, all of which can vary importantly among the objects. Since narrow + broad component fits are not possible for all individual galaxies, the identification relied largely on visual inspection. Following G14, we classified each galaxy as having a secure (unambiguous presence), a candidate (possible or marginal), or no detection of a broad component around the H$\alpha$+[NII] complex. Weaker and/or lower velocity outflows may be missed; the outflow incidences based on this identification procedure may thus represent lower limits.

### 2.5.4. Binning and LOESS Representations of 2D Distributions

The size and homogeneous coverage over the physical properties explored enabled us to split the objects in several tens of bins sampling the parameter space. More specifically, in our finest grids, we split the full sample in about 60 bins containing each about 10 galaxies. To recover and visualize the mean trends in various pairs of galaxy properties, we found the non-parametric locally-weighted polynomial regression method LOESS, as implemented by Cappellari et al. (2013)[5], particularly useful. We assigned the properties derived from the binned data to the individual galaxies, and accounted for Poisson uncertainties. We typically employed a second-order polynomial and a 0.5 fraction of points in the local approximation, and validated the recovered trends against the input binned distributions.

Figure 3 illustrates the above steps, showing the incidence of a broad outflow emission component as a function of stellar mass and MS offset. We determined the presence of the broad component from the spectra of individual galaxies, distinguishing between secure and candidate cases as explained in the previous subsection (left panel of Figure 3). We defined seven bins in log($M_*$) of width between 0.3 and 0.5 dex, with ~100 galaxies in each of the central five bins and about 30 in the lowest and highest mass bins. For every mass bin, we then defined $\Delta$MS intervals containing ~10 galaxies each, with typical width of 0.15 dex but varying from ~0.05 dex close to the MS at intermediate masses and up to ~1 – 3 dex for the lowest $\Delta$MS bin below the MS. We then computed the fraction of galaxies with a broad outflow component (and its Poisson uncertainty) for each of the 61 resulting bins, assigning a weight of 1 and 0.5 to secure and candidate detections, respectively (middle panel of Figure 3). Specifically, this fraction is defined as $f_{\rm out}$ = ($N_{\rm secure}$+0.5$\times N_{\rm candidate}$)/$N_{\rm total}$, where $N_{\rm total}$ is the number of galaxies in the bin[6]. These fractions are used as input for the two-dimensional LOESS smoothing, with the resulting distribution highlighting the main underlying trends (right panel of Figure 3). Because of the modest number of galaxies in each bin, the relative uncertainties can be large especially for low incidence bins (exceeding ~50% for $f_{\rm out}$ ≲ 0.20), but are taken into account in the LOESS smoothing. We note that all trends in outflow incidence presented in this work remain qualitatively the same if we exclude the candidates (or weight them as secure detections), which reflects the fairly similar distributions of candidate and secure cases in the galaxy parameters explored.

## 3. RESULTS

### 3.1. Incidence of Broad Emission Components with Stellar Mass and Star Formation Rate Properties

The identification of broad emission in individual galaxies (Section 2.5.3) yielded 190 objects with significant or tentative outflow signature, for a global fraction of 32%; 117 of them have a secure detection (20%). Their distributions in stellar mass versus MS offset, and the binned and LOESS representations (Figure 3) clearly indicate a strong increase of incidence of broad components from low mass, low ΔMS to high mass, high ΔMS. The overall mass trend is clearly dominant; the incidence correlates strongly with the median log($M_*$) in the bins with Spearman rank correlation coefficient of ρ=0.66 and 5.1σ significance, and more weakly with median ΔMS with ρ=0.30 at the 2.3σ level. Below the MS into the quiescent regime, outflows are detected in a comparable fraction of galaxies as near the MS at similar mass but the statistical uncertainties are larger; the sparser sampling and smaller number of galaxies in this region of parameter space make an assessment of trends more difficult.

---
[5] We used the IDL routine CAP_LOESS_2D of Cappellari et al. (2013) that implements the multivariate LOESS algorithm of Cleveland & Devlin (1988), available at http://www-astro.physics.ox.ac.uk/~mxc/software/#loess.

[6] Throughout the paper, fractions of objects of given galaxy parameters are computed relative to the total number of galaxies from the full sample with the same parameters unless explicitly stated otherwise.

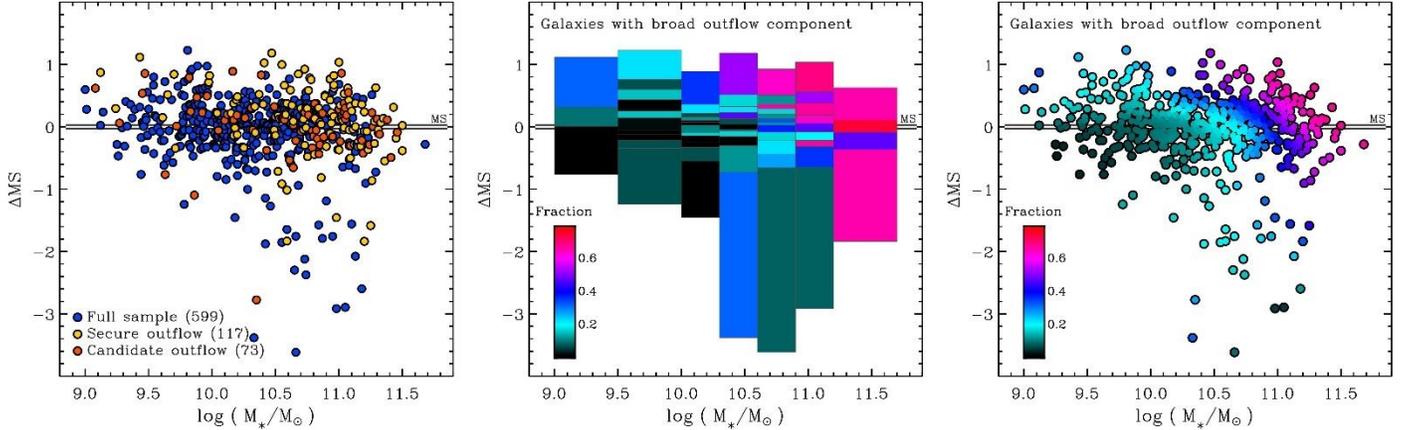

**Figure 3.** Distribution of the incidence of broad emission components associated with ionized gas outflows as a function of stellar mass and offset in SFR from the MS. The broad component is identified in the individual galaxy spectra around Hα+[NII], and the different representations in the three panels are created as described in Sections 2.5 and 3.1. *Left:* Distribution of individual galaxies of the full sample in log($M_*$) vs ΔMS, color-coded by whether they exhibit a secure, a candidate, or no broad component in their spectrum around Hα+[NII] (yellow, orange, and blue, respectively). *Middle:* Fraction of galaxies with a broad component in different bins of log($M_*$) and ΔMS, color-coded according to the color bar in the lower left of the panel. *Right:* Trend in parameter space after LOESS smoothing based on the binned data, sampled at the position of each galaxy in log($M_*$) vs ΔMS, color-coded with the broad component incidence as shown with the lower-left color bar. Fractions were computed assigning a weight of 1 and 0.5 to secure and candidate outflow detection, respectively.

In another approach, we fitted to the unweighted averaged Hα+[NII] spectra in each bin a combination of three narrow components (of equal width) and a single broad component, and determined the fraction of broad flux to narrow Hα flux. The choice of a single broad component here enables us to assess whether it is present or not down to low amplitudes where fits with three Gaussians are insufficiently constrained (see Section 2.5.3). The results are shown in Figure 4, using the LOESS-smoothed representation. The range is here restricted to ΔMS > −1.25 dex because of the sparser sampling at lower ΔMS and some of the objects exhibit fairly strong [NII] emission that affects the fitted parameters of the single broad component[7]. This contamination makes the ratio of $F_{br}$ with $F(Hα)_{na}$ a less meaningful and more uncertain measure of the outflow incidence in this case. Within ΔMS ± 1.25 dex where a comparison is most reliable, the flux-based method yields qualitatively similar trends to those obtained from the incidence-based approach in Figure 3. The relative strength of the broad component increases by a factor of 10−20 from the low- to high-mass end of the full sample. A more modest increase by a factor of ∼2 is detected in ΔMS at fixed mass from the lower to the upper tail of the MS, such that the combined trend is comparable to that of the incidence of broad components seen in Figure 3.

These two approaches consistently show the very rapid rise in outflow incidence (and strength) above the Schechter mass. Seventy to eighty percent of the most massive galaxies in our sample exhibit outflows, confirming and strengthening the results of G14, and decreasing the typical (mean and median) uncertainties in the incidence to ±0.1 per log($M_*$)−ΔMS bin and ±0.05 per mass bin.

---

[7] The fit results for the full set of 61 bins are shown in Figure 8, and the trend of higher [NII]/Hα$_{br}$ with higher [NII]/Hα$_{na}$ found in our data is discussed in Section 4.2.1.

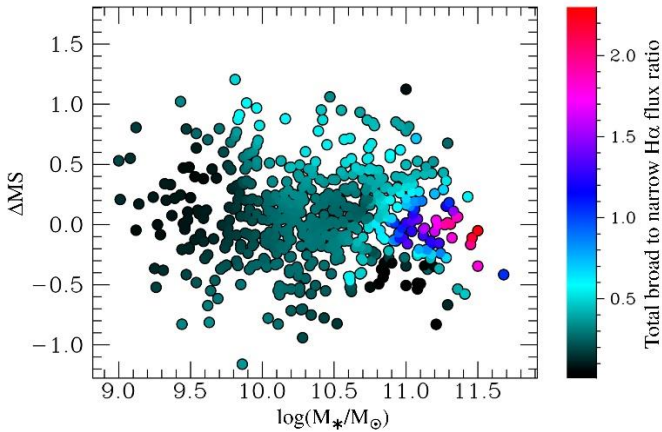

**Figure 4.** Distribution of the total broad to narrow H$\alpha$ flux ratio as a function of stellar mass and MS offset. The flux ratio was obtained by fitting a narrow component to each of H$\alpha$ and the [NII] doublet lines, and a single broad underlying component to the (unweighted) average spectrum of galaxies in bins of log(M$_*$) vs $\Delta$MS. The bins are the same as shown in Figure 3 but restricted to $\Delta$MS $> -1.25$ dex as explained in Section 3.1. The data points represent individual galaxies, color-coded by the LOESS-smoothed flux ratio, following the methodology described in Section 2.5.4. The importance of the broad emission increases strongly with stellar mass and also, though more modestly, from below to above the MS. These trends are qualitatively similar to those derived from the incidence of a broad component in the individual spectra plotted in Figure 3.

*3.2. Separation into SF-driven and AGN-driven Outflows*

We discussed in Section 1 the evidence for both SF-driven and AGN-driven outflows in high-redshift SFGs. N12a, N12b, FS14, and G14 have shown that the broad emission from SF-driven outflows originates from regions that are spatially extended across the entire star-forming disk (on $\sim 4-10$ kpc scales), above a "break-out" threshold of $\Sigma_{\rm SFR} \sim 1$ M$_\odot$ yr$^{-1}$ kpc$^{-2}$, at all galaxy masses (see also Davies et al. 2019). The broad emission is visible in H$\alpha$, [NII], and [SII], and has a typical FWHM line width of $\sim 400-500$ km s$^{-1}$. In contrast, the broad emission from AGN-driven outflows is centrally concentrated, with a FWHM extent of $\sim 1-3$ kpc when resolved in AO-assisted IFU data or strongly lensed objects. Such outflows are associated with massive galaxies hosting prominent bulges, are characterized by larger line widths of FWHM $\sim 1000-2000$ km s$^{-1}$, and typically exhibit narrow [NII]/H$\alpha$ flux ratios $> 0.45$ (hereafter "[NII]-strong"), in the AGN/LINER region of rest-optical diagnostic diagrams (e.g., Baldwin et al. 1981; Veilleux & Osterbrock 1987; Kewley et al. 2001, 2013; Kauffmann et al. 2003). The broad emission from AGN-driven outflows is dominated by [NII], with [NII]/H$\alpha_{\rm br} \sim 1-2.5$ that is significantly higher than for SF-driven winds. In simpler fits assuming a *single* broad component as in the previous subsection, the strong broad [NII]$\lambda$6584 emission obviously leads to a redshift of a few hundred km s$^{-1}$ relative to the narrow H$\alpha$ line (an effect also seen in some local AGNs; e.g., Ho et al. 1997).

Outflows driven by star formation or by AGNs can thus be readily distinguished by their spectral properties, in both narrow and broad components. Figure 5 illustrates the distinction for our sample with the weighted averages of the highest S/N spectra of galaxies with SF-driven and AGN-driven outflows (including 33 and 30 sources, respectively) after applying the classification scheme described below. The stack for AGN-driven outflows exhibits much larger velocity widths in the broad emission components, and higher [NII]/H$\alpha$ ratios in both narrow and broad components.

*3.2.1. AGN Identification*

Because the strength of the broad outflow emission varies with galaxy properties, and because most of the data were obtained in natural seeing conditions, it is not possible to distinguish between SF- and AGN-driven outflows in all galaxies based on the above spatial and spectral characteristics. Instead, we used as discriminant the narrow component [NII]/H$\alpha$ flux ratio[8] together with diagnostics from ancillary data from X-ray to mid-IR and radio wavelengths. Specifically, galaxies with [NII]/H$\alpha_{\rm na} > 0.45$ and/or X-ray, mid-IR, or radio properties indicative of an AGN are identified as hosting an AGN. These different diagnostics are known to select partly different subsets of the full AGN population notably because of the variability and phenomenology of AGN activity (e.g., Juneau et al. 2013; Coil et al. 2015; Azadi et al. 2017; Padovani et al. 2017), motivating our approach of using complementary indicators. When present, the broad H$\alpha$+[NII] component is then attributed to an AGN-driven outflow if the galaxy has an AGN, or to SF-driven outflows otherwise.

For the AGN identification from the ancillary data, we followed a similar approach as described by G14 with updated source catalogs, based on the combination of the following diagnostics: (1) X-ray detection and X-ray-based properties (Xue et al. 2011; Symeonidis et al. 2014); (2) mid-IR *Spitzer*/IRAC 5.8$\mu$m$-$3.6$\mu$m vs 8$\mu$m$-$4.5$\mu$m colors (Donley et al. 2012); (3) 1.4 GHz "radio excess threshold" comparing the measured flux to that expected from the SFR (e.g., Appleton et al. 2004; Delvecchio et al. 2017); (4) detection in wide-field VLBI observations; (5) a match with optically or X-ray variable sources. The list of surveys and catalogs used is given in Appendix A. Since the sensitivity, instruments used, and availability of the observations vary considerably between the different extragalactic fields relevant to our full sample, the identification from ancillary data is not uniform and likely misses a higher proportion of AGN in shallower fields.

---

[8] For the cases of weak broad emission in individual spectra, the [NII]/H$\alpha$ ratio from fits assuming a single Gaussian for each line is used as proxy, as motivated by the simulations of Förster Schreiber et al. (2018; Appendix C).

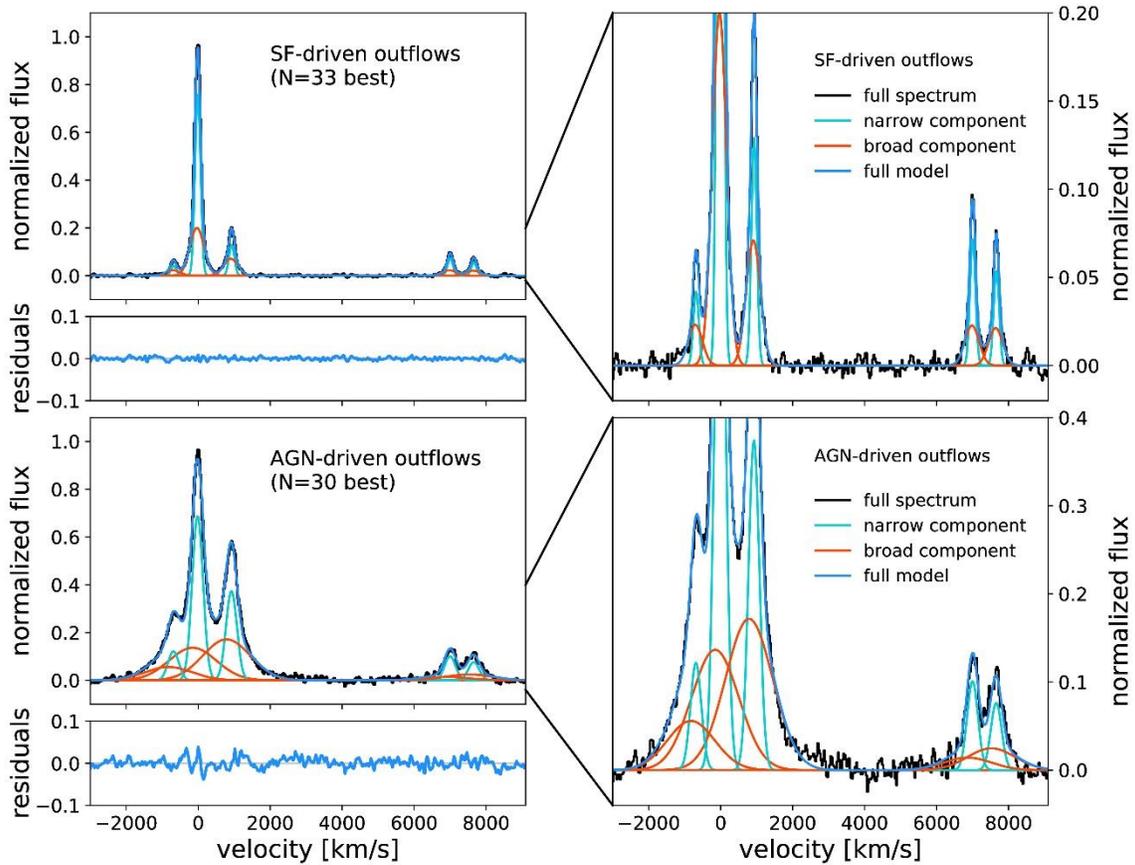

**Figure 5.** Weighted averages of the highest S/N spectra showing the distinction between broad emission associated with SF- and AGN-driven outflows. *Top panels:* Stack for 33 galaxies with SF-driven outflows. *Bottom panels:* Stack for 30 galaxies with AGN-driven outflows. In each of the large panels, the stacked spectrum (black solid line) is plotted as a function of velocity relative to the peak of H$\alpha$. The best-fit from multiple narrow and broad Gaussian profiles to the H$\alpha$, [NII], and [SII] lines (following the MCMC fitting described in Section 2.5) is shown: the total narrow + broad emission (light blue curves), the narrow component (cyan curves), and the broad component (orange curves). The residual spectrum is plotted below each stacked spectrum on the left-hand side. The right-hand side panels show the same spectra as on the left, zoomed in on a smaller flux range. The good quality detection of the broad [SII] emission in the stack for SF-driven outflows (FWHM$_{br}$ ~ 460 km s$^{-1}$) enables a robust determination of the electron density of the outflowing gas (Section 4.1). The detection of broad [NII] and [SII] emission for the AGN-driven outflows rules out a dominant BLR origin and implies an unbound nuclear outflow with FWHM$_{br}$ ~ 1550 km s$^{-1}$. The narrow and broad component [NII]$\lambda$6584/H$\alpha$ ratios are 0.54 and 1.26, respectively, characteristic of AGN/LINER excitation. Given their different characteristics, emission from outflows driven by star formation or by AGNs can be readily distinguished from the spectral properties.

Ideally, AGN identification from rest-optical line emission would require at least one pair of ratios such as the classical "BPT" diagnostic [OIII]$\lambda$5007/H$\beta$ vs [NII]$\lambda$6584/H$\alpha$ (e.g., Baldwin et al. 1981; Veilleux & Osterbrock 1987; Kauffmann et al. 2003; Kewley et al. 2006, 2013). Measurements of all four lines are available only for a small number of objects in our sample (discussed by Kriek et al. 2007, Newman et al. 2014, FS14, and G14), so we relied on [NII]/H$\alpha$ for most galaxies. A ratio above ~0.45 indicates a contribution by a non-stellar excitation source, which we attributed to AGN activity. This [NII]/H$\alpha$ cut may miss some AGN in low-metallicity hosts towards lower stellar masses (Kewley et al. 2013). In addition, nebular line emission from star-forming regions within the apertures used to extract the spectra may lower the overall line ratio sufficiently (a well-known effect, e.g., Ho et al. 1997; FS14; Coil et al. 2015); weaker AGN in actively star-forming systems can be difficult to detect, especially in seeing-limited data of distant galaxies.

Among the full sample, 574 galaxies have relevant data in the X-ray, mid-IR, or radio ranges and 70 of them (12%) satisfy at least one of the corresponding AGN criteria. X-ray-identified AGN dominate, with 54 sources. From the [NII]/H$\alpha_{na}$ ratio, 119 of all 599 galaxies (20%) are identified as having an AGN. In addition, six objects show evidence for very broad emission (FWHM $\gtrsim$ 3000 km s$^{-1}$) in H$\alpha$ but not in the [NII] and [SII] forbidden lines and for a bright, central point-like source in their HST rest-UV/optical imaging. The latter properties are consistent with Type 1 AGN dominated by emission from the unobscured broad line region (BLR) in the close vicinity of the nucleus; five of the six BLR sources are also identified as AGN from the ancillary data. In total, there are thus 152 AGN (25%) in the full sample, of which 146 are associated with obscured Type 2 AGN (24%); 38 objects fulfil both the X-ray/mid-IR/radio and [NII]/H$\alpha_{na}$ sets of criteria. Not all AGN identified from the ancillary data exhibit AGN signatures in their rest-optical spectra; this is the case for 32 sources in our sample. Conversely, 81 AGN are identified solely from their elevated [NII]/H$\alpha_{na}$ ratio.

The global fractions of AGN in our full sample are consistent with those of AGN surveys at $z \sim 1 - 3$ using similar AGN indicators (e.g., Reddy et al. 2005; Daddi et al. 2007; Kriek et al. 2007; Brusa et al. 2009; Xue et al. 2010; Aird et al. 2012; Bongiorno et al. 2012; Hainline et al. 2012; Juneau et al. 2013; Coil et al. 2015; Wang et al. 2017; Padovani et al. 2017). With the average S/N ~ 15 per spectral channel of our galaxy spectra (Section 2.5.1), [NII]/H$\alpha$ ratios of 0.45 are measured with an uncertainty lower than 10% (and ratios down to ~0.1 are determined with a 3$\sigma$ significance); measurement uncertainties are thus unlikely to dominate the partial cross-identification between diagnostics. The AGN fraction in flux limited surveys increases as a function of galaxy stellar mass, and different diagnostics select partly complementary subsets of AGN (see references above). Our use of combined diagnostics and the fact that our sample is weighted towards high-mass galaxies explains that our fractions tend to lie at the higher end of ranges reported in the literature when considering the full mass range. The non-uniform depth of the ancillary data between the various fields is also reflected in the corresponding AGN fractions. In particular, the fraction of X-ray-identified AGN decreases by a factor of 2.6 between the deepest GOODS-South field and the shallower COSMOS field (from 13% to 5%), emphasizing the importance of using complementary techniques.

The AGN among our sample span a wide range of bolometric AGN luminosities $\log(L_{AGN}/[\text{erg s}^{-1}]) \sim 42.5 - 47$. We estimated the $L_{AGN}$ for the X-ray identified AGN from the published X-ray fluxes (corrected for H absorption in most cases; see references in Appendix A), applying the bolometric correction to the derived hard 2-10 keV luminosity of Rosario et al. (2012). For the AGN identified from [NII]/H$\alpha_{na}$, we inferred the [NII]$\lambda$6584 AGN luminosity from the narrow component H$\alpha$ flux and [NII]/H$\alpha_{na}$ ratio. Since the apertures used to extract the spectra encompass regions at least a few kpc in diameter, we accounted for a likely contribution by star-forming regions to the narrow [NII] emission based on the (evolving) mass-metallicity relation as parametrized by Genzel et al. (2015), and the conversion from log(O/H) to [NII]/H$\alpha$ of Pettini et al. (2004)[9]. The [NII] luminosity was scaled to the bolometric luminosity assuming a fiducial [NII]$\lambda$6584/[OIII]$\lambda$5007 ~ 0.75 (the mean and median for bright Seyfert 2 galaxies in the SDSS survey) and a bolometric conversion based on Netzer et al. (2009). For AGN hosts satisfying both the X-ray and [NII]/H$\alpha_{na}$ criteria, the respective $\log(L_{AGN})$ estimates are in broad agreement, with a median difference of ~0.5 dex and a scatter of ~0.8 dex.

All these estimates have large uncertainties but are sufficient for an order-of-magnitude assessment. The distributions of X-ray and [NII]-based $\log(L_{AGN})$ largely overlap, with median values of 44.7 and 45.2, respectively. Given the large uncertainties, this difference is not significant but we note that the very deep X-ray data in the GOODS-South field probe AGN down to the lowest luminosities, and our strict [NII]/H$\alpha_{na}$ cut would miss weaker AGN in galaxies towards lower masses and metallicities, and with stronger outshining from star-forming regions. The specific AGN luminosity together with the black hole to galaxy stellar mass ratio $M_{BH}/M_*$ provides a measure of the Eddington ratio, $\log(\lambda_{Edd}) = \log(L_{AGN}) - 38.1 - \log(M_{BH})$. Assuming a universal $M_{BH}/M_* = 0.0014$ (Häring & Rix 2004), the AGN in our sample have a broad distribution spanning ~4.5 dex and peaking at $\log(\lambda_{Edd}) \sim -1$, with no obvious differentiation between the subsets identified from X-ray and [NII]/H$\alpha_{na}$. This peak value is broadly consistent with the break in the Eddington ratio distribution of X-ray-selected AGN at $z \sim$

---

[9] Assuming all of the narrow [NII] emission is excited by the AGN increases the estimates by an overall factor of ~2.

1 − 3 (e.g., Aird et al. 2012; 2018; Hickox et al. 2014; Bernhard et al. 2018). In summary, the AGN identified in our sample span a *broad range* from low luminosity, very sub-Eddington AGNs (log($\lambda_{Edd}$) ~ −3.5) to luminous QSOs with high Eddington ratios (with 10 objects, or 7%, at log($\lambda_{Edd}$) ≳ 0), with distribution peaks around the typical values as inferred from X-ray-selected surveys.

### 3.2.2. Trends of Incidence of SF- and AGN-driven Outflows in Stellar Mass and Star Formation Properties

With the above AGN identification, the broad emission component is attributed to AGN-driven winds in 103 of the 190 galaxies with outflow signatures, and to SF-driven winds in the other 87 galaxies. There are 43 AGN in which no outflow signature is detected. Given the shortcomings noted above for our [NII]/H$\alpha_{na}$ plus ancillary data-based classification, it is possible that in some objects the dominant driver of the outflow is misidentified. It is also possible that AGN- and SF-driven outflows coexist in the same source, as best seen in a few of the SINS/zC-SINF targets with high-resolution AO data (FS14). However, the spectral differences seen in Figure 5, and the distinction in the various trends and physical properties of the AGN- and SF-driven outflows discussed below, suggest that "cross-contamination" is not important.

The incidence of AGN and AGN-driven outflows, denoted $f_{AGN}$ and $f_{AGNout}$, is shown in Figure 6, in the log($M_*$) vs ∆MS plane, and also as a function of log($M_*$) for wider ∆MS bins below, around, and above the MS. The incidence of AGN and AGN-driven nuclear outflows appears to be solely a function of mass, with almost two orders of magnitude increase between log($M_*$/M$_\odot$) ~ 9 and > 11. The correlation with stellar mass is very strong (ρ=0.84 for AGN and ρ=0.81 for AGN-driven outflows for the binned data, both significant at ≈ 6.5σ). AGN and AGN-driven winds are not correlated with ∆MS (ρ=−0.10 and 0.04, respectively). At the highest masses, up to ~80−100% of the galaxies harbour an AGN, and most (~60−75%) drive a prominent outflow. This is equally true above, on, and below the MS. We note that because our AGN identification procedure includes the [NII]/H$\alpha_{na}$ criterion in addition to the X-ray/mid-IR/radio indicators, our resulting $f_{AGN}$ are higher than reported by G14. Counting only AGN identified based on the ancillary indicators, the $f_{AGN}$ for the present sample are ~2 − 3 times lower, and at most ~40–55% at the highest masses, similar to the fractions in G14.

Other near-IR studies also report high fractions ~50-75% of AGN-driven outflows among X-ray-selected QSOs and more moderate luminosity AGNs based on [OIII]λ5007 and/or Hα+[NII] kinematic signatures (e.g., Brusa et al. 2015b; Harrison et al. 2016). On the other hand, Leung et al. (2017) found a lower fraction of 19% among z ~ 2 AGNs identified from X-ray, IR, and rest-optical indicators, observed as part of the MOSDEF survey. Since AGNs identified by various diagnostics form a subset of the entire galaxy population, part of the differences in reported fractions may be attributed to the different sample selection. Our main goal of characterizing the role of outflows among the overall galaxy population motivated our analysis of a sample selected irrespective of the nuclear activity of the galaxies. In our sample and over all masses (excluding BLRs), outflows are detected in 103 of the 146 AGNs for a global fraction of 71%, or 60% when weighting outflow candidates by ×0.5. There is no obvious distinction between AGNs identified through different diagnostics. For instance, 37 of the 49 AGN sources detected in hard X-ray emission (irrespective of their [NII]/H$\alpha_{na}$ ratio) exhibit an outflow signature, and 90 of the 119 AGN galaxies identified based on the [NII]/H$\alpha_{na}$ criterion (irrespective of their X-ray properties) do, for nearly equal unweighted fractions of 76% or ~ 60% when downweighting candidates. The stacked spectra of these subsets of AGNs with outflows are very similar to each other and to the stack from all AGN-driven outflows.

The dependence of AGN or AGN-driven outflow incidences is at least quadratic in stellar mass, and possibly exponential, with a sharp onset around log($M_*$/M$_\odot$) ~ 10.7−10.9. This threshold coincides with the Schechter mass, independently of redshift across the z = 0.6 − 2.7 range our data sample. Again, our results are in excellent agreement with, and further strengthen the conclusions of G14. Since the Schechter mass corresponds to the transition above which the likelihood of quenching rises strongly at all redshifts z ≲ 2 − 3, thus limiting galaxy growth (Peng et al. 2010b; Ilbert et al. 2013; Muzzin et al. 2013), the threshold for the onset of prominent AGN-driven nuclear outflows appears to be concomitant to quenching. This point is discussed further in Section 4.

The incidence of outflows driven by star formation, $f_{SFout}$, i.e., in galaxies without indication of AGN activity from either the [NII]/H$\alpha_{na}$ or the ancillary data indicators, is shown in Figure 7 (left panel). The $f_{SFout}$ does not depend on stellar mass (ρ = −0.01) but increases upward with ∆MS at all masses, with ρ = 0.50 (3.9σ) in the binned data. About 25−30% of "starbursting outliers" above the MS (∆MS ≳ 0.6 dex) drive a SF-driven outflow detected in rest-optical line emission. Physically, one would expect that the incidence of (detectable) SF-driven outflows also depends on the SFR surface density, $\Sigma_{SFR}$ (e.g., Heckman 2002; Kornei et al. 2012; N12b; G14). Figure 7 (right panel) shows the distribution in the $\log(M_*)$ vs $\log(\Sigma_{SFR})$ plane, where the surface density is taken as half the total SFR uniformly distributed within $R_e$. The incidence increases with $\Sigma_{SFR}$, with significant fractions $\gtrsim 10\%$ above $\sim 0.5$–1 $M_\odot$ yr$^{-1}$ kpc$^{-2}$, corresponding to the observed threshold above which the broad emission signature becomes strong (e.g., N12b; Davies et al. 2019).

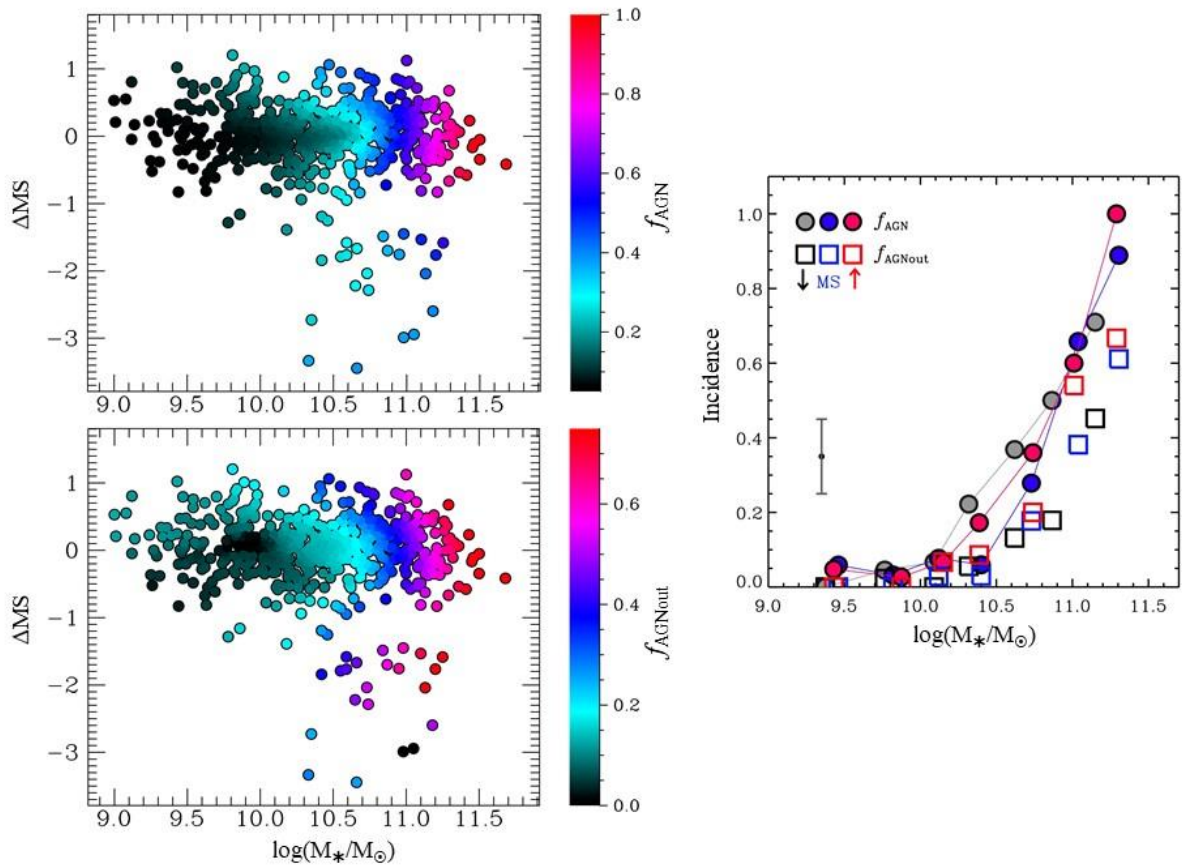

**Figure 6.** Distribution of the incidence of AGN and AGN-driven outflows with stellar mass and MS offset. *Left:* the top panel shows the LOESS smoothed representation for the incidence of galaxies hosting an AGN, based on identification through X-ray/mid-IR/radio properties and the [NII]/Hα narrow component criterion, as described in Section 3.2.1. The bottom panel shows the LOESS smoothed representation for the incidence of galaxies with AGN-driven outflow. The color-coding follows the color bars and is adjusted for each plot to the respective mininum to maximum values, in order to emphasize the trends in each quantity. *Right:* Variation of incidence for the full sample, now binned in three ΔMS intervals: below (black/grey symbols), on (blue symbols), and above (red symbols) the MS. The incidence of AGN and AGN-driven outflows is plotted with filled circles and open squares, respectively, and the average uncertainty is shown by the vertical error bar. The incidence of outflows and AGNs (or outflow properties such as velocity widths and broad-to-narrow line flux ratios) do not significantly depend on redshift (G14 and Section 3.3) such that we marginalize over this parameter throughout the paper. The incidence of both AGN and AGN-driven outflows correlates strongly with stellar mass irrespectively of location relative to the MS (Section 3.2.2), and exhibits a steep onset around the Schechter mass at $\log(M_*/M_\odot) \sim 10.8$.

Galactic winds identified from interstellar absorption features in rest-frame UV spectra of SFGs at $z \sim 0.5$–3 are more prevalent ($\gtrsim 50\%$; e.g., Steidel et al. 2010; Weiner et al. 2009; Kornei et al. 2012; Martin et al. 2012; Rubin et al. 2014). Our lower detection rate could be due to S/N limitations in individual spectra (since the broad SF-driven outflow signature typically has a modest amplitude and velocity width; Figure 5), different sample selection, or possibly reflects the different outflow phase probed by each technique. The Hα+[NII] emission line technique is sensitive to the emission measure and thus probes preferentially ongoing ejection of denser gas. The rest-UV absorption line technique integrates over the line-of-sight and down to more tenuous material, and would thus more easily detect outflows even of low duty cycle at any given time. These factors may also explain differences in outflow incidence trends with star formation properties, which are found to be typically weak or absent based on rest-UV interstellar absorption tracers (e.g., Weiner et al. 2009; Kornei et al. 2012; Martin et al. 2012; Rubin et al. 2014). Observations of sizeable samples targeting both interstellar absorption and nebular line tracers of outflows in the same galaxies would be valuable to understand these differences.

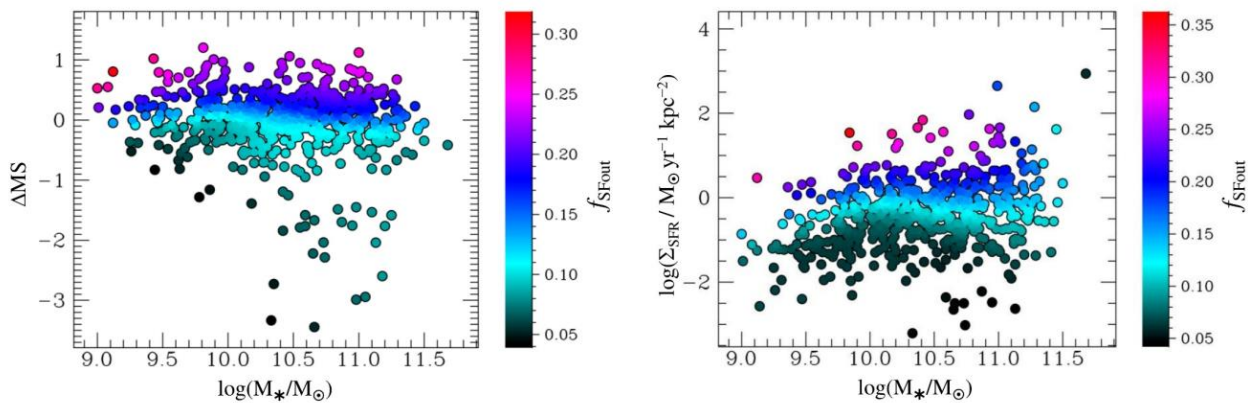

**Figure 7.** LOESS-smoothed distribution of the incidence of SF-driven outflows with stellar mass and star formation properties. *Left:* Incidence in the log($M_*$) vs MS offset $\Delta$MS. *Right:* Incidence in the log($M_*$) vs SFR surface density, $\Sigma_{SFR}$. The color coding is adjusted to cover the range between minimum and maximum values in each plot separately, as shown by the color bars. The outflow incidence and properties do not significantly depend on redshift (N12b; G14; and Section 3.3) such that we marginalize over this parameter throughout the paper. The incidence of SF-driven outflows correlates with both $\Delta$MS and $\Sigma_{SFR}$ at all masses, with no trend as a function of log($M_*$) (Section 3.2.2).

### 3.2.3. Extended Trend Analysis

We investigated correlations between larger sets of parameters to include spectral properties and additional galaxy parameters. We quantified trends between pairs of properties with the Spearman rank correlation coefficient and examined all properties simultaneously through a Principal Component Analysis (PCA). Because the coverage in galaxy parameters of the sample varies somewhat with redshift and some trends may not be monotonic, we validated the main trends by inspecting the 2D distributions in properties (similarly to Figure 3) and controlling for specific variables where appropriate. In particular, the low-mass coverage is different for the galaxies at $0.6<z<1.2$, $1.2<z<1.8$, $1.8<z<2.7$, leading to some spurious correlations with redshift. In G14, we found that the broad outflow emission incidence and spectral properties did not depend significantly on redshift when splitting the sample in two $z$ intervals. With the larger sample analyzed here, no significant redshift dependence at fixed galaxy property is found either (see also Table 1). For the trend analysis below, we thus marginalized over redshift.

In a first step, we considered variations in the binned log($M_*$)–$\Delta$MS plane of the incidences $f_{SFout}$, $f_{AGNout}$, $f_{AGN}$, the [NII]/H$\alpha_{na}$ and $F_{br}/F(H\alpha)_{na}$ flux ratios, and the broad component velocity width FWHM$_{br}$. The spectral properties were derived from fits to the H$\alpha$+[NII] complex in the unweighted stacks, with a single component for the broad emission in order to characterize the emission across all bins, as described in Section 2.5.3. Figure 8 (left six panels) shows the fractions and spectral properties plotted at the average log($M_*$) and $\Delta$MS values of the galaxies in each bin (using the median instead makes little difference in the distributions and derived trends). In addition to the trends for the incidences discussed in the previous subsection, [NII]/H$\alpha_{na}$ and $F_{br}/F(H\alpha)_{na}$ correlate most strongly with log($M_*$) ($\rho \sim 0.76$ at $5.9\sigma$, and $\rho \sim 0.47$ at $3.6\sigma$) and little with $\Delta$MS ($\rho < 0.15$). The FWHM$_{br}$ correlates primarily though moderately with log($M_*$) ($\rho \sim 0.4$ at $\sim 3\sigma$).

These trends, and the strong associations among subsets of properties, are most clearly visualized in the results from the PCA shown in the rightmost panel of Figure 8. For these data and eight parameters, 71% of the total sample variance is explained by two principal components (PCs), with PC 1 and 2 strongly coupled to log($M_*$) and $\Delta$MS, respectively. The most striking feature of the loading plot is the nearly orthogonal separation between $f_{SFout}$ on one hand, and $f_{AGNout}$, $f_{AGN}$ and spectral characteristics on the other, implying in particular that SF- and AGN-driven outflows are strongly decoupled from each other in their incidence and emission properties. The association between $f_{AGN}$, [NII]/H$\alpha_{na}$, and log($M_*$) is not surprising given the AGN identification procedure of Section 3.2.1 and the galaxy stellar mass-metallicity relationship (e.g., Erb et al. 2006; Queyrel et al. 2009; Zahid et al. 2014; Wuyts et al. 2014, 2016a; Sanders et al. 2015, 2018; Kashino et al. 2017). To some extent, that with FWHM$_{br}$ and $F_{br}/F(H\alpha)_{na}$ may be influenced by a more important contribution to the line profile from steep but unresolved inner velocity gradients in the more massive and denser galaxies (or residuals stemming from uncertainties in the velocity field used to remove the orbital motions). However, tests using model data cubes of massive rotating disks with a bulge, subjected to the typical seeing and noise level of the data, and analyzed in the same way indicate that such effects could only account for a line FWHM of $\sim$ 400 km s$^{-1}$ (G14). This value is well below the FWHM $\gtrsim$ 1000 km s$^{-1}$ measured in the high-mass bins.

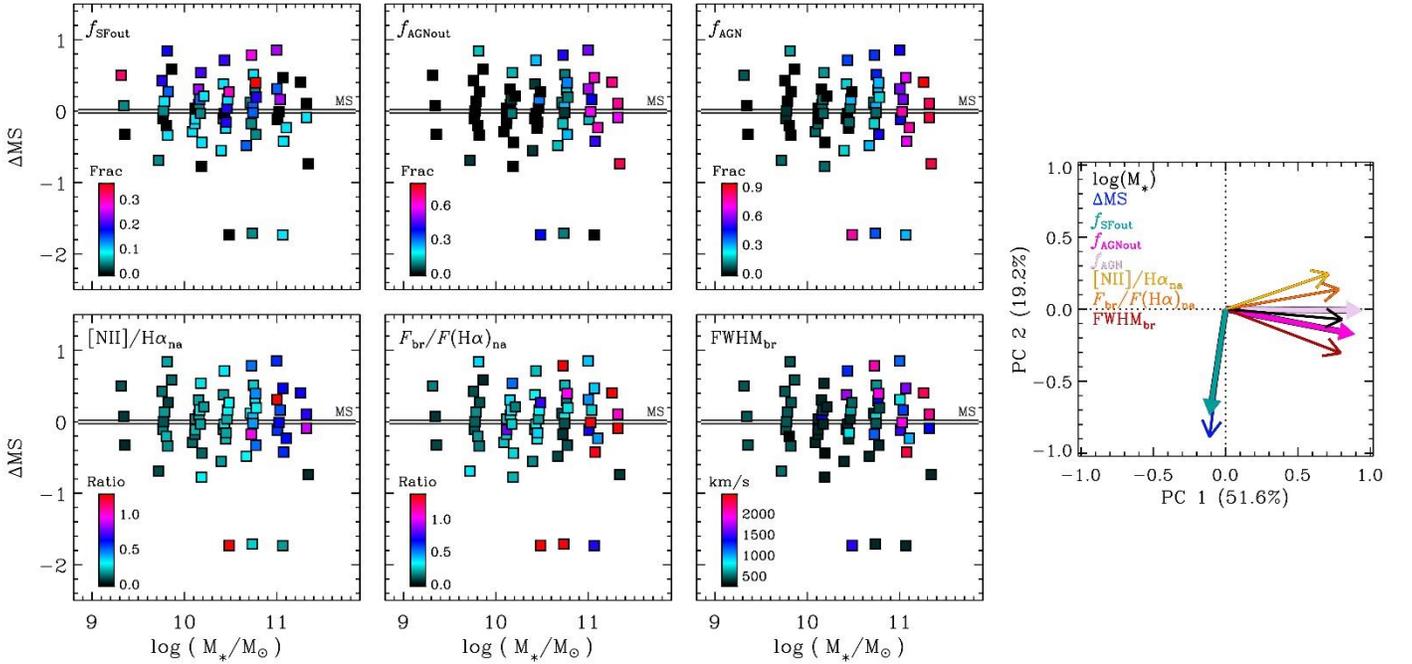

**Figure 8.** Distribution of incidence and spectral properties in the log($M_*$)–$\Delta$MS plane, and results of the PCA on the data. Left panels: The top three panels show the fraction of galaxies with a SF-driven outflow, an AGN-driven outflow, and an AGN in bins of stellar mass and MS offset; the bottom three panels show the [NII]/H$\alpha$ narrow component flux ratio, the broad to H$\alpha$-narrow flux ratio, and the velocity FWHM of the broad emission from fits to the unweighted stacked spectra with a single Gaussian encompassing the entire broad emission around the H$\alpha$+[NII] complex (see Section 2.5.2). The data points are plotted at the average log($M_*$) and $\Delta$MS of the galaxies in each bin, and color-coded to represent the third property according to the color bar shown in each panel. The color coding is adjusted independently for each panel to cover the corresponding range between the minimum and maximum values. The horizontal line indicates the MS in the normalized vertical units. Right panel: Projection of the properties onto the first two principal components (PCs) that account for 71% of the total variance of the data; the arrows show the loadings for the different properties color-coded according to the legend in the top left corner. PC 1 and 2 are strongly associated with log($M_*$) and $\Delta$MS, respectively. The incidence of SF-driven outflows is tightly correlated with $\Delta$MS, while that of AGN-driven outflows and AGN and the spectral properties are strongly correlated with log($M_*$).

Focussing on the incidences, we searched for additional significant trends with a larger set of galaxy parameters. We considered the stellar mass surface density $\Sigma_*$ within $R_e$, the absolute and specific star formations rates SFR and sSFR, and $\Sigma_{SFR}$ within $R_e$. We also included the structural parameters $\Delta MR_e$, $R_e$, the projected major-to-minor axis ratio $q$ and the Sérsic index $n_{\mathrm{Sérsic}}$ from the HST $H$-band imaging (van der Wel et al. 2012; Lang et al. 2014). We computed the incidences rebinning the full sample in each of these galaxy properties paired with stellar mass following the same procedure as described in 2.5.2, with about 60 bins of $\sim$10 galaxies each. We also carried out a PCA using individual galaxy parameters, and replacing fractions for the incidence with categorical variables representing the absence or presence of an outflow or AGN with 0 and 1, respectively, and a candidate outflow with 0.5.

This more comprehensive exploration shows that $f_{AGNout}$ and $f_{AGN}$ correlate roughly equally with $\Sigma_*$, and $\Sigma_{*,1\mathrm{kpc}}$ ($\rho \sim 0.8$ at the $\sim$6$\sigma$ level) as with log($M_*$), and also depend significantly on $n_{\rm Sérsic}$ and $\Delta MR_e$, which are related to compactness ($\rho \sim 0.4$–$0.5$ at the $\sim 3$–$4\sigma$ level). The $f_{\rm SFout}$ correlates most tightly and in similar measure with $\Delta MS$ and $\log(sSFR)$ ($\rho \sim 0.5$ at the $\sim 4\sigma$ level), and is somewhat more loosely coupled to SFR and $\Sigma_{\rm SFR}$ ($\rho \sim 0.4$ at the $\sim 3\sigma$ level). The PCA indicates that the first PC is largely a measure of central stellar mass concentration (largest positive loading from $\Sigma_*$, largest negative loading from $\Delta MR_e$), the second PC is related to SF activity and intensity (large positive loadings from all of $\Delta MS$, $\log(SFR)$, $\log(sSFR)$, $\log(\Sigma_{\rm SFR})$), and the third PC is most closely connected to size properties (largest loading from $R_e$). These three PCs account for 71% of the total variance. There is no significant correlation of outflow incidence with $q$ for our sample, suggesting that inclination may not play an important role in terms of the detectability of ionized gas outflows in $z \sim 0.6 - 2.7$ galaxies.

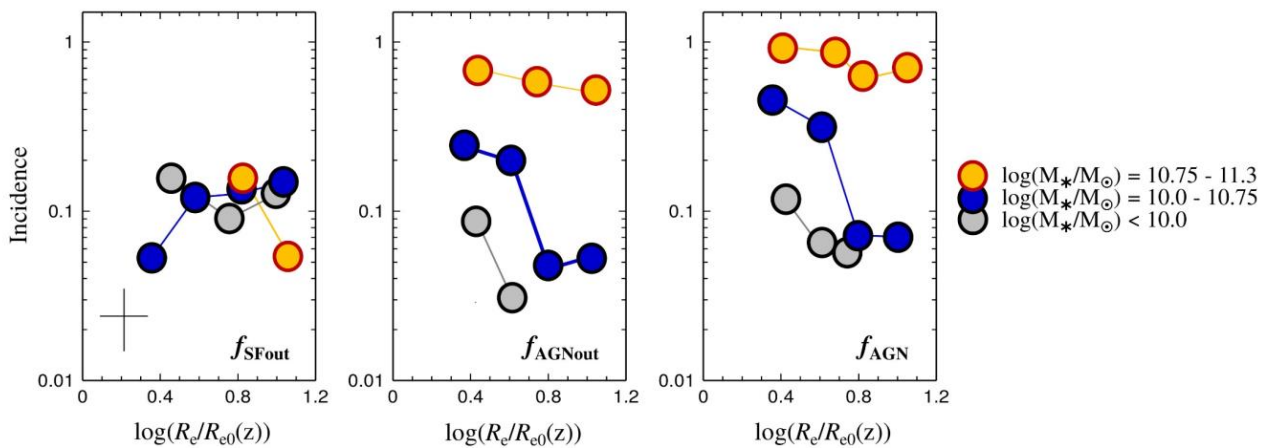

**Figure 9.** Incidence of outflows and AGN as a function of effective radius corrected for the mean size evolution of SFGs ($\langle R_{e0} \rangle \propto (1+z)^{-0.75}$; van der Wel et al. 2014). *Left:* Incidence of SF-driven outflows. *Middle:* Incidence of AGN-driven outflows. *Right:* Incidence of AGN. The galaxies are split in two mass intervals below the Schechter mass ($\log(M_*/M_\odot) < 10$ and $10 < \log(M_*/M_\odot) < 10.75$, plotted as grey and blue circles, respectively) and one interval above it ($\log(M_*/M_\odot) > 10.75$, yellow circles). The typical uncertainty is shown in the lower left corner of the left-hand panel. The incidence of SF-driven outflows at all masses does not depend significantly on size. For AGN-driven outflows, as well as for AGN, the incidence also does not vary significantly on size above the Schechter mass but it anticorrelates at lower masses.

The trends between $f_{\rm AGNout}$ and $f_{\rm AGN}$ with central stellar mass density and galaxy size are consistent with the elevated AGN fraction among compact SFGs in $\log(M_*/M_\odot) \gtrsim 10$ samples (e.g., Barro et al. 2014; Rangel et al. 2014; Kocevski et al. 2017; Wisnioski et al. 2018). Most interestingly, perhaps, the anticorrelation with $R_e$ and the correlation with $\Sigma_*$ in our sample are clearly stronger in the mid- to low-mass regime and significantly weaker above the Schechter mass. Figure 9 plots the incidences as a function of $R_e$, normalized to the mean redshift evolution in size for SFGs ($\propto (1+z)^{-0.75}$; van der Wel et al. 2014), for two mass intervals below and one above $\log(M_*/M_\odot) = 10.75$. Given the typical uncertainties, the anti-correlation with relative size of $f_{\rm AGNout}$ and $f_{\rm AGN}$ is only significant below the Schechter mass. In contrast, $f_{\rm SFout}$ shows no significant trend in all mass ranges.

In summary, the formal analysis of the new sample of 599 galaxies robustly confirms the main trends reported in our earlier N12, FS14 and G14 work based on much smaller samples. The incidence of AGN-driven outflows (and AGN) depends primarily and strongly on galaxy stellar mass and central surface density, while that of SF-driven outflows depends primarily on star formation properties. They are almost completely decoupled from each other in the $\log(M_*) - \Delta MS$ plane. Over the best sampled region of parameter space, the outflow velocities and the broad-to-narrow flux ratio are dominated by the mass trend. After accounting for the cosmic evolution in galaxy properties (such as the MS zero-point), the incidence of outflows and AGN, and the outflow emission properties, do not significantly depend on redshift.

## 4. DISCUSSION

In this Section, we estimate the physical characteristics of the outflows, and explore trends thereof, based on the rest-optical emission line properties. This analysis relies on line profile decomposition of stacked spectra with a broad and narrow component for each of the H$\alpha$, [NII] doublet, and [SII] doublet lines (as described in Section 2.5.3). Given the typical strength of the broad emission, it is necessary to consider wider bins in various properties to optimize the S/N and ensure that the fitted parameters are sufficiently well constrained. Table 1 gives the key parameters from the line fitting, along with average properties of the galaxies entering each stack, and derived physical quantities. The Table also includes the circular velocity $v_c$ at $R_e$ derived from the (narrow component) H$\alpha$ kinematics, in the disk framework and accounting for beam-smearing and the effects of pressure support[10] (e.g., Burkert et al. 2016; Wuyts et al. 2016b; Genzel et al. 2017; Übler et al. 2017; FS18). The outflow velocity $v_{out}$ was determined by taking the full width at zero power (FWZP) of the entire H$\alpha$+[NII] complex, subtracting 1600 km s$^{-1}$ (the velocity separation between the [NII] doublet lines), and dividing by two. Local electron densities were estimated from the [SII]$\lambda$6716/$\lambda$6731 ratios (using the calibration of Sanders et al. 2016).

The other properties were derived following closely Genzel et al. (2011), N12, and G14, to which we refer for details. We summarize here the method and main assumptions. To compute the mass outflow rate $\dot{M}_{out}$, we assumed that the gas is outflowing in a cone of solid angle $\Omega$, with a radially constant $\dot{M}_{out}$ and $v_{out}$. These assumptions are motivated by observations of the dependence of MgII absorber occurrence and profiles as a function of inclination of the host galaxy (e.g., Bordoloi et al. 2011; Kacprzak et al. 2012; Bouché et al. 2012) as well as theoretical work on both energy- and momentum-driven outflows (e.g., Veilleux et al. 2005; Murray et al. 2005; Hopkins et al. 2012). We further assumed that the gas is photoionized, and in case B recombination with an electron temperature of $T_4 = T_e/10^4$ K $= 1$. In this simple model, the average electron density and volume filling factor of the outflowing ionized gas scale with radius as $R^{-2}$ (for a constant mass outflow rate) but the local electron density $<n_e^2>^{1/2}$ of filaments or compact clouds from which the broad H$\alpha$ emission component originates does not vary significantly with radius. For case B recombination, the effective volume emissivity is $\gamma_{H\alpha}(T) = 3.56\times 10^{-25} T_4^{-0.91}$ erg cm$^{-3}$s$^{-1}$ (Osterbrock 1989). The total ionized gas mass outflow rate, independent of $\Omega$, can then be obtained from the extinction-corrected, optically thin H$\alpha$ luminosity $L(H\alpha)_{0,br}$ via

$$L(H\alpha)_{0,br} = \gamma_{H\alpha}(T) \int \Omega R^2 n_e(R) n_p(R) dR,$$

$$M_{HII,He} = \mu \cdot \int \Omega R^2 n_p dR = \frac{\mu L(H\alpha)_{0,br}}{\gamma_{H\alpha}(T) \times n_e}, \text{ and}$$

$$\dot{M}_{out} = \Omega R^2 \mu\, n_p(R) v_{out} = M_{HII,He} \cdot \frac{v_{out}}{R_{out}} \quad (2).$$

Here, $n_p$ is the proton density, $\mu = 1.36 \times m_p$ is the effective nucleon mass for a 10% helium fraction, $M_{HII,He}$ is the mass in ionized H and in He, and $R_{out}$ is the outer radius of the outflow. High-resolution SINFONI+AO data for a few tens of our sample galaxies indicate that the ionized gas outflows have sizes of at least $\sim 2-3$ kpc, and are typically extended over the central half-light radius (N12a,b; FS14); we thus adopted $R_{out} = R_e$.

The intrinsic H$\alpha$ luminosity of the broad component $L(H\alpha)_{0,br}$ was computed by scaling the average narrow component extinction-corrected H$\alpha$ luminosity $L(H\alpha)_{0,na}$ of galaxies contributing to a stacked spectrum according to the broad-to-narrow H$\alpha$ ratio $F_{br}/F_{na}$. We derived the $L(H\alpha)_{0,na}$ from the observed fluxes corrected for dust attenuation, adopting the visual extinction towards the bulk of stellar light $A_{V,stars}$ from the best-fit SED models (Section 2.3), using the Calzetti reddening curve, and accounting for extra attenuation towards the nebular gas with $A_{gas}=A_{stars}\times(1.9-0.15\times A_{stars})$ (Wuyts et al. 2013; see also Price et al. 2014). As discussed by Genzel et al. (2011), this estimate likely provides a conservative lower limit to $L(H\alpha)_{0,br}$ because the blueshift of the broad H$\alpha$ profile relative to the narrow H$\alpha$ emission in several individual cases and in stacked spectra (by $\sim 30-140$ km s$^{-1}$) suggests a significant amount of differential extinction within the outflowing component. The intrinsic narrow component H$\alpha$ luminosity was used to derive the SFR within the same (projected) regions where broad outflow emission is detected via the Kennicutt (1998) conversion adjusted to our adopted Chabrier (2003) IMF. From these quantities, we then calculated the outflow momentum rate $\dot{p}_{out} = \dot{M}_{out} \times v_{out}$ relative to the radiation momentum rate from the star-forming population $\dot{p}_{rad} = L_{SFR}/c$, and the outflow kinetic energy rate $\dot{E}_{out} = \frac{1}{2}\dot{M}_{out} \times v_{out}^2$ relative to the luminosity from the young stars $L_{SFR}$, with $L_{SFR}/L_\odot \sim 10^{10} \times$ SFR/M$_\odot$ yr$^{-1}$ (e.g., Kennicutt 1998).

---

[10] Spatially-resolved mapping from near-IR IFU observations have demonstrated that at least ~70% of z ~1 – 3 SFGs have disk-like kinematics (e.g., Förster Schreiber et al. 2009, 2018; Épinat et al. 2009, 2012; Wisnioski et al. 2015, 2018; Stott et al. 2016; Tiley et al. 2016; Harrison et al. 2017).

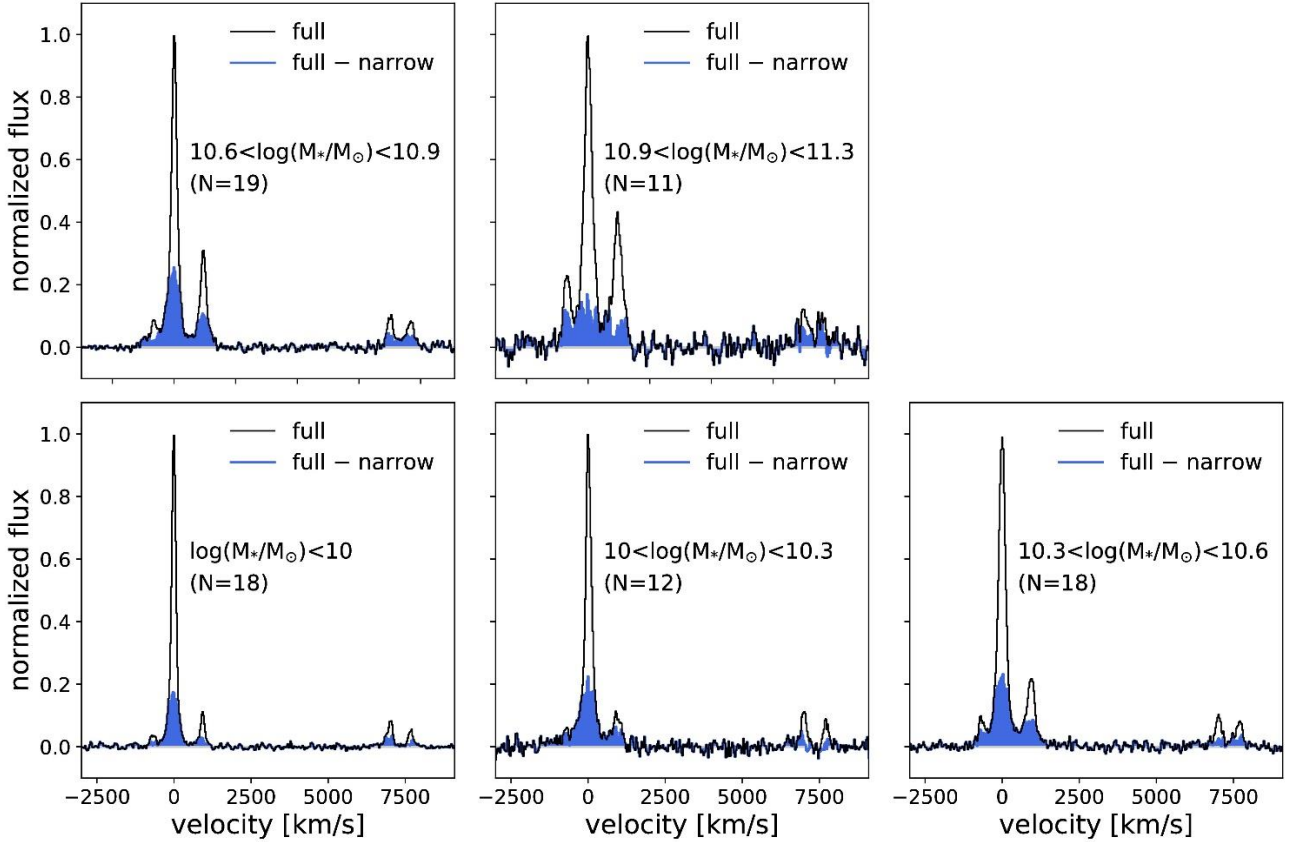

**Figure 10.** Stacked spectra of galaxies with SF-driven outflows in five bins of stellar mass (with 11−19 SFGs per bin, including secure outflow detections and candidates). *Top row:* Weighted averaged spectra in the two highest mass bins. *Bottom row:* Weighted averaged spectra in the three lowest mass bins. In each panel, the full stacked spectrum is plotted with the black line; the broad emission component obtained after subtracting the best-fit narrow emission to Hα and the [NII] doublet is shown in blue.

*4.1. Star Formation-Driven Outflows*

We focus first on the outflows driven by star formation, i.e., in galaxies that show no indication of AGN activity in the X-ray, mid-IR, or radio regime and have [NII]/Hα$_{na}$ < 0.45. Figure 10 shows the stellar mass dependence of the broad emission component profile, from stacked spectra of galaxies with an individual outflow detection (including candidates, at half the weight) in five stellar mass bins. To best highlight the broad emission, the residual spectra after subtraction of the best-fit narrow component in Hα and [NII] are plotted alongside the full stacked spectra. We used the spectral fits to explore possible trends of the mass loading factor on galaxy mass, and the energetics of the SF-driven outflows at $z$~1–3. The results are given in Table 1.

*4.1.1. Mass Dependence of Mass Loading*

Figure 11 presents the mass dependence of the inferred outflow mass loading factor and of the ratio of outflow velocity to galaxy circular velocity. Except for the lowest mass bin, $v_{out}$ is approximately constant while $v_c$ increases from ~120 km s$^{-1}$ at the lowest mass to ~305 km s$^{-1}$ at the highest mass. Consequently, $v_{out}/v_c$ decreases with mass, from values >2 in the two lower mass bins to ≲1.5 in the two higher mass bins. Given the uncertainties in these estimates, this trend is tentative. However, it would be consistent with SF-driven winds in the lower mass objects escaping the galaxies more easily and perhaps penetrating into the outer halo while the higher mass SFGs can only drive fountains, as has been proposed theoretically for some time (e.g., Dekel & Silk 1986; Murray et al. 2005; Oppenheimer & Davé 2008; Übler et al. 2014).

In momentum- or energy-driven winds, one would expect that the outflow mass loading factor depends on galaxy mass as follows:

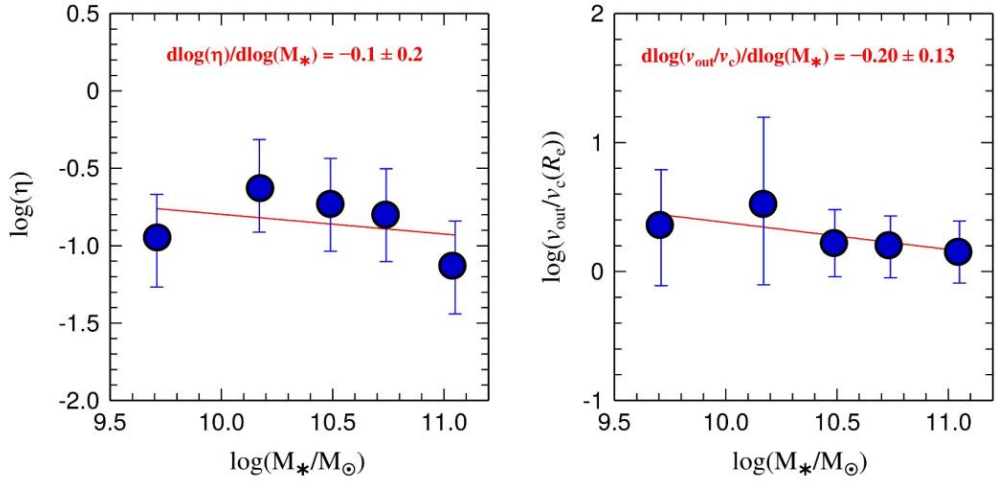

**Figure 11.** Stellar mass dependence of SF-driven outflow properties. *Left*: Mass loading factor versus $\log(M_*)$. *Right*: Ratio of outflow velocity to galaxy circular velocity at $R_e$. Each data point represents a stack of 11-19 SFGs and the red line in each panel shows the best-fit slope to the measurements. Given the shallow slopes, and within the uncertainties, the data are consistent with little, if any, dependence of mass loading factor with galaxy mass. The trend in $v_{\rm out}/v_{\rm circ}$ suggests the SF-driven outflows may escape from the low-mass galaxies but probably drive fountains instead at higher masses.

$$\dot{M}_{out} v_{out} = L_{SFR}/c \propto SFR \sim M_*  \quad \text{on the MS,} \quad (3),$$

$$\text{or } \dot{M}_{out} v_{out}^2 \propto L_{SFR} \propto SFR, \quad \text{such that}$$

$$\eta = \frac{\dot{M}_{out}}{SFR} \propto v_{out}^{-1} \text{ (momentum) or } v_{out}^{-2} \text{ (energy).}$$

If $v_{\rm out} = a \times v_c$ (e.g., Oppenheimer & Davé 2006) and $v_c \propto (M_*/m_d)^{1/3}$ (e.g., Mo et al. 1998), where $m_d$ is the fraction of the central galaxy relative to that of the entire halo, then $\eta \propto M_*^\alpha$, with $\alpha$ in the range $-1/3$ to $-2/3$.

Over the roughly 1.5 dex range in average $\log(M_*)$ for the stacked spectra, this scaling would predict a drop in mass loading by a factor of ~3 for momentum-driven winds, and ~10 for energy-driven winds. Assuming a constant local electron density in the outflowing ionized gas of $n_{\rm e,br} = 380$ cm$^{-3}$, the mass loading factor inferred empirically from our observations is fairly constant across this mass range, with a slope in logarithmic units of $-0.1 \pm 0.2$ (Figure 11). This $n_{\rm e,br}$ value is most robustly constrained from the best-fit [SII]$\lambda6731/\lambda6716_{\rm br} = 1.07 \pm 0.13$ for the stack of the 33 highest S/N spectra of galaxies with secure SF-driven outflow detection (Table 1 and Figure 5). We note that a nearly identical value of $1.08 \pm 0.14$ is obtained without prior on the narrow [SII] ratio[11]. This measurement represents the first somewhat significant (~2.5$\sigma$ away from the low-density limit) determination of the local electron density in the broad component in the literature. For comparison, the lower S/N stack of 14 SFGs of N12b yielded a 2$\sigma$ limit of <1200 cm$^{-3}$, with a ratio consistent with the low-density limit. Since $\eta$ is inversely proportional to $n_{\rm e,br}$, a variation with mass could still occur if the electron density of the outflowing ionized gas changes substantially with stellar mass. Because the [SII] lines are weak, S/N considerations hamper a reliable narrow+broad component decomposition in the stacked spectra for five mass bins, but splitting in two mass bins does not indicate that $n_{\rm e,br}$ changes significantly with mass.

Our finding of a roughly constant, and fairly low, mass loading factor over the $\log(M_*)$ range probed is surprising in view of the slope of the observed relationships between galaxy metallicity, $M_*$, and SFR as well as the mass dependence of $M_*/M_{\rm halo}$ from abundance matching studies up to $\log(M_*/M_\odot) \sim 10.7$, which appear to require a power-law slope of $\alpha \sim -0.35$ to $-0.8$, with $\eta \sim 0.3 - 1$ or higher around $\log(M_*/M_\odot) \sim 10$ (Lilly & Carollo 2013; Davé et al. 2017). In their analysis of outflow properties in stacked slit spectra of ~ 125 non-AGN SFGs at z ~ 2 from the MOSDEF survey, Freeman et al. (2017) found a positive correlation between $\eta$ and $\log(M_*)$, which also is discrepant with theoretical expectations. Since H$\alpha$+[NII] observations probe ~$10^4$ K ionized gas, it is conceivable that an important part of the expelled mass is contained in other outflow phases. Nonetheless, the indication of lower $v_{out}/v_c$ at higher masses from our data could be consistent with more massive galaxies having higher baryon mass fractions and metallicities by retaining their baryons more efficiently than lower mass galaxies, at least up to $\log(M_*/M_\odot) \sim 10.7$.

---

[11] This fit with the narrow [SII] ratio as free parameter yields [SII]$\lambda6731/\lambda6716_{\rm na} = 1.33 \pm 0.09$, fully consistent with the prior of $1.34 \pm 0.03$.

### 4.1.2. Mass and Energetics of the Winds

Keeping in mind the substantial uncertainties of the above estimates and the simplicity of our model, resulting in systematic uncertainties of the mass ejection, momentum, and energy rates by at least ± 0.3 dex, the inferred outflow rates are typically ~10 $M_\odot$ yr$^{-1}$ (and in the range ~ 0.2 – 20 $M_\odot$ yr$^{-1}$) and the mass loading factors are $\eta \sim 0.1 - 0.2$ with our adopted $n_{e,br} = 380$ cm$^{-3}$. These mass loading factors are ~3–10 times lower than in our earlier work (Genzel et al. 2011; N12a,b; FS14; G14; see also Erb et al. 2006), owing to the higher $n_{e,br}$ estimate based on the present data. As noted above, previous [SII] ratio measurements for the broad component were poorly constrained, and previous estimates for the mass loading factors assumed $n_{e,br}$ ~50–80 cm$^{-3}$. Our new results imply that the outflowing gas has on average both a higher local electron density and a higher [NII]/Hα ratio than the narrow emission component (Figures 5 and 10). These findings suggest that the broad component-emitting gas may be compressed by shocks.

The modest mass loading factors, along with the ratios of outflow to stellar radiation momentum rates ~ 0.1 – 1 and energy rates ~ 10$^{-4}$ – 10$^{-3}$, can be easily accounted for by single scattering, photon pressure in momentum-driven winds (e.g., Murray et al. 2005; Hopkins et al. 2011). However, these low mass loading factors would not seem to be sufficient to account for the low galaxy formation efficiencies inferred at low masses from abundance matching studies (Behroozi et al. 2013a,b; Moster et al. 2013, 2018). We stress again that all our derived outflow properties pertain to the ~10$^4$ K ionized gas phase; any additional contribution from very hot ionized plasma or cold atomic and molecular material in the outflows would increase the mass outflow rates and mass loading factors. The total mass loading factors could be much higher if the warm ionized phase of the SF-driven outflows carries only a small fraction of the total mass (as well as momentum and energy) of the winds, as inferred for local starbursts driving winds where the neutral and cold molecular phases dominate the mass outflow rates and the hot phase dominates the energetics (e.g., Veilleux et al. 2005; Heckman & Thompson 2017; and references therein).

### 4.2. AGN-Driven Outflows

#### 4.2.1. The [NII]/Hα Excitation Sequence

We now turn to AGN-driven outflows. One of the distinctive spectral feature of the galaxies hosting AGN-driven winds is the high excitation implied by their elevated [NII]/Hα ratios. The good statistics afforded by our sample at high stellar masses (with a total of 119 [NII]-strong sources) enabled us to investigate properties as a function of [NII]/Hα$_{na}$ ratio. Above the mass threshold where AGN-driven outflows (and AGN) become common (i.e., at $\log(M_*/M_\odot) \gtrsim 10.6$), [NII]-strong sources with a wide range of [NII]/Hα$_{na}$ ratios (0.45 to 5.6) are stochastically distributed in $\log(M_*)$ vs ΔMS space. Figure 12 shows the stacked spectra of the objects split in five bins of [NII]/Hα$_{na}$ ratio. It is apparent that as the *narrow* [NII]/Hα ratio increases, so does the *broad* [NII]/Hα. Overall, the [NII]/Hα$_{br}$ line ratio is typically between 1 and 2, indicating that the outflowing component likely is dominated by shocks. Here again, a reliable broad+narrow profile decomposition for [SII] is hampered by the combination of the weakness of the [SII] lines and S/N, and is further complicated by the strong blending resulting from the faster winds. Splitting instead the AGN-driven outflows in two bins of low and high [NII]/Hα$_{na}$ ratio or redshift, or considering the stack of 30 of the highest S/N spectra (Figure 5, bottom), yields higher local electron densities $n_{e,br}$ ~ 1000 cm$^{-3}$ for the broad emission compared to the narrow emission, and also compared to the densities obtained for the SF-driven outflows. Comparably high electron densities ~1000 cm$^{-3}$ have been inferred from rest-optical line ratios tracing warm ionized gas in integrated spectra of local AGNs (e.g., Perna et al. 2017). Despite the uncertainties, the trends in line excitation suggest that compressing shocks are common in the outflowing components of AGN-driven outflows (e.g., Kewley et al. 2010; 2013).

In addition to the line excitation properties, a key difference for the AGN-driven outflows is their higher velocities compared to the SF-driven outflows. The $v_{out}/v_c$ ratios are $\gtrsim$ 3.5 for the various stacks considered (see Table 1) and up to about 8 for the "best stack," systematically above the range of ~1.5−3.5 for the SF-driven winds. Thus, in contrast to the SF-driven outflows, the AGN-launched outflows can in principle escape the galaxy and even its halo. In a collision with gas coming in from outside the halo, the shocked wind will attain a post-shock temperature of ~5.5×10$^7$ K, which has a very long cooling time, and will thus remain as a hot atmosphere that could prevent gas from entering the galaxy (e.g., Bower et al. 2017).

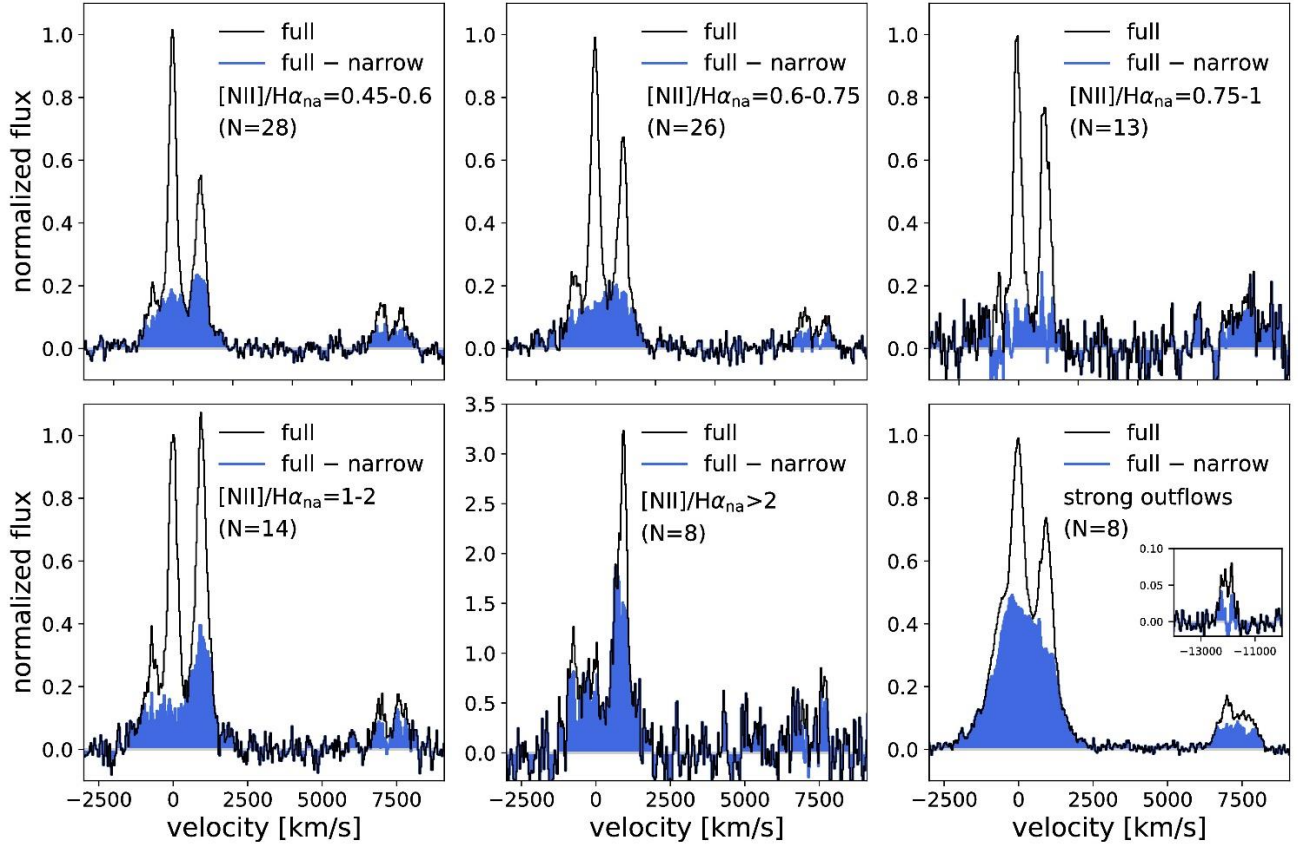

**Figure 12**. Stacked spectra of galaxies with AGN-driven outflows in five bins of narrow component [NII]/Hα ratio (with 8−28 objects per bin, including secure outflow detections and candidates). *Top row:* Weighted averaged spectra in the three lowest ratio bins. *Bottom row, left and middle:* Weighted averaged spectra in the two highest ratio bins. *Bottom right:* Weighted stack of the eight objects with particularly strong broad-to-narrow flux ratio, which also exhibits [OI]λ6300 line emission (inset). As in Figure 10, the full stacked spectrum is plotted in black and the broad emission component after subtraction of the best-fit narrow emission is shown in blue.

Nine objects further stand out in their spectral properties by their particularly strong broad emission component, indicative of more powerful outflows (we refer to them as "strong outflows"). The stack of the eight objects whose spectra encompass the [OI]λ6300 emission line is plotted in Figure 12 (bottom right), from which a high $F_{br}/F_{na} = 1.3$ is derived. In the stack, as well as in some of the individual spectra, [OI] is well detected. The total [NII]λ6584/Hα, [SII]λ6716+6731/Hα, and [OI]λ6300/Hα ratios (0.7, 0.25, 0.05, respectively) are consistent with a mixture of stellar photoionization, hard radiation from the AGN, and shocks contributing to the line excitation (e.g., Kewley et al. 2006; Allen et al. 2008; Rich et al. 2010, 2011; see also G14). These nine "strong outflows" sources are systematically more actively star-forming[12] and more compact than the other galaxies hosting an AGN-driven outflows in the same $\log(M_*/M_\odot) > 10.7$ range, with median $\Delta MS = +0.63$ dex and $\Delta MR_e = -0.25$ dex compared to −0.01 and −0.11 dex, respectively. They also appear to have a more elevated nuclear accretion activity, with a median $\log(L_{AGN}/[\text{erg s}^{-1}]) \sim 45.7$ and $\log(\lambda_{Edd}) \sim -0.5$ compared to ∼ 45.4 and −1.0 for the other similarly massive objects with an AGN-driven outflow. Possibly, both star formation and AGN activity contribute importantly to driving the observed winds in these "strong outflows" systems. Despite their above-average outflow mass ejection rates (∼ 110 $M_\odot$ yr$^{-1}$), and momentum and energy injection rates, the higher average SFR implies $\eta \sim 0.45$, $\dot{p}_{out}/\dot{p}_{rad} \sim 10$, and $\dot{E}_{out}/L_{SFR} \sim 0.02$. The latter estimates are comparable to the other AGN-driven outflows (Table 1), which are eight times more frequent in our sample. These properties are consistent with enhanced outflows being associated with increased AGN and star formation activity in absolute terms, but the scarcity of these objects suggests that this phase is rather short-lived and does not dominate in a population- and time-averaged sense.

---

[12] All but one of these objects have a SFR estimate based on a Herschel/PACS 160 μm detection, with negligible contamination from AGN emission (e.g., Rosario et al. 2012).

Summarizing, for the majority of the AGN-driven outflows, and adopting a common $n_{e,br} \sim 1000$ cm$^{-3}$, the inferred mass outflow rates and mass loading factors are in the ranges $\sim 5 - 45$ M$_\odot$ yr$^{-1}$ and $\sim 0.1 - 0.5$, respectively. These estimates are comparable or somewhat higher than those of the SF-driven outflows but because of the larger wind velocities, they carry ~10 times greater momentum, and ~50 times greater energy. If the density in the outflow were lower, the differences would be even larger. Overall, the AGN-driven outflows have typical inferred $\dot{p}_{out}/\dot{p}_{rad} \sim 1 - 15$, and $\dot{E}_{out}/L_{SFR} \sim 0.002 - 0.06$, supporting that the AGN is required to power the observed winds. The ratio of AGN to stellar bolometric luminosities estimated for these galaxies spans a very wide range, but is typically $L_{AGN}/L_{SF} \sim 1$, such that the momentum and energy ratios remain roughly the same when referred to the AGN properties. The high ratios suggest the AGN-driven outflows may be energy-driven (e.g., Fabian 2012; Zubovas & King 2012; Faucher-Giguère & Quataert 2012).

### 4.3. Connection Between AGN Outflows and Activity, Star Formation, and Gas Content

Our most important finding of Section 3 and by G14 was that incidence and average strength of AGN-driven outflows depend essentially only on stellar mass, and not or very little on redshift and on star formation properties. The sharp increase in $f_{AGNout}$, as well as $f_{AGN}$ can be described by a combination of exponential functions parametrized in terms of $x=\log(M_*)$ as g(x) = a·exp((x−x$_0$)/0.2)/(1+exp((x-x$_0$)/0.2), with the critical mass around the Schechter mass $x_0 \sim \log(M_S/M_\odot) \sim 10.8$. The coincidence of the sharp onset of powerful high-velocity outflows with the stellar mass describing the characteristic "mass quenching" scale (e.g., Peng et al. 2010) is very interesting, but does it imply a causal connection (e.g., Lilly & Carollo 2016)?

If the AGN-driven outflows are causally connected to mass quenching, then their incidence should also (anti-)correlate with specific SFR and molecular gas mass fraction. Figure 13 compares the incidence of AGN-driven outflows, and of AGN, near the MS to the relationships for sSFR (in the parametrization of Whitaker et al. 2014) and for molecular gas fraction $\mu = M_{molgas}/M_*$ (from the scaling relations of Tacconi et al. 2018; their Table 3b). This Figure shows an anti-correlation in the sense and in the mass regimes expected although this broad coincidence alone may not suffice to establish a causal link. We note that we inversed the direction of the vertical axis for sSFR and $\mu$ to turn the trends into an apparent correlation between plotted quantities, to highlight the similarity in their behaviour.

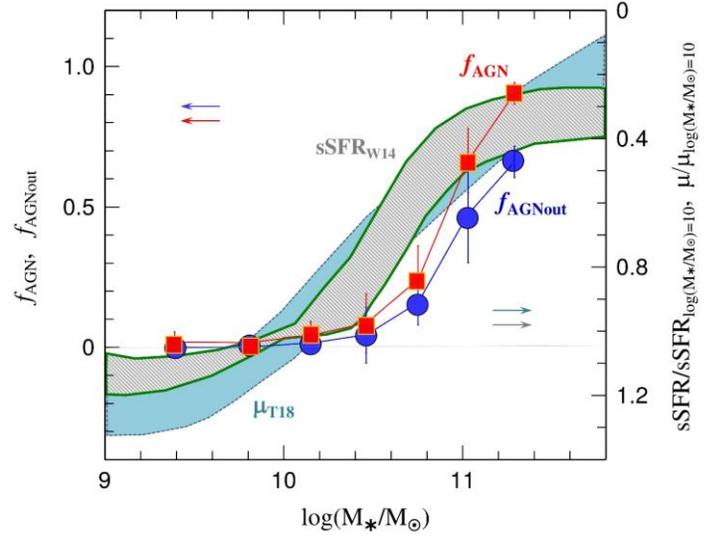

**Figure 13.** Variation with stellar mass of the incidence of AGN activity and outflows, of star formation rate, and of molecular gas content for the region $\Delta$MS ±0.4 dex around the MS of z = 0.7 – 2.6 SFGs. The fractions of galaxies identified as hosting an AGN, and as driving an AGN-driven outflow, are plotted with red squares and blue circles, respectively, on the left-hand side vertical scale. The relationship for the specific SFR (sSFR) of Whitaker et al. (2014) is drawn as the grey-shaded region with green boundary, and that for the molecular gas mass fraction $\mu = M_{molgas}/M_*$ of Tacconi et al. (2018) is plotted as cyan-shaded region, both normalized at $\log(M_*/M_\odot) = 10$ and scaled according to the vertical axis on the right side of the panel. The width of the shaded regions corresponds to the dispersion around the relationships. The sSFR and $\mu$ scale is inversed to ease the comparison of the trends with mass between all four quantities, and to highlight the similarity in location and shape of the rapid transition region near the Schechter mass.

### 4.4. The Role of Outflows in Galaxy Evolution

Combining the results from this and the previous Sections, we find that SF-driven and AGN-driven outflows during the peak epoch of galaxy star formation and supermassive black hole accretion 6 – 11 billion years ago appear to operate in two distinct regimes of the stellar mass vs SFR space. SF-driven winds occur mainly at high levels of star formation activity above the MS, at all masses, when the momentum and energy injected by supernovae, radiation, and stellar winds presumably can overcome the self-gravity of the gas layer (e.g., N12b). The moderate and apparently mass-independent outflow velocities suggest that at $\log(M_*/M_\odot) > 10.3$, these winds cannot escape the potential of the galaxies but instead drive fountains, as expected theoretically (e.g., Dekel & Silk 1986; Oppenheimer & Davé 2008; Davé et al. 2011; Übler et al 2014). To be efficient in driving gas out in the general equilibrium baryon cycle framework (e.g., Bouché et al. 2010; Lilly et al. 2013), the mass loading factors of outflows has to be near or above unity (e.g., Hopkins et al. 2014; Muratov et al. 2015). If the low mass loading factors of the warm ionized component inferred from our data are generally applicable, the majority of the momentum or energy of the winds would have to be in a hot, or a cold molecular phase.

As galaxies grow along the MS, the $M_{BH} - \sigma_*$ relation between black hole mass and stellar velocity dispersion (e.g., Kormendy & Ho 2013) implies that their central massive black holes should on average grow in lockstep. If so, the Bondi-Hoyle accretion rate from the vicinity of the black hole increases with the square of the black hole's mass, thus leading to a run-away increase in accretion, potentially triggering much increased AGN activity and nuclear AGN-driven winds. The fast nuclear winds, possibly acting even in periods when the Eddington ratio is moderate or low (Pillepich et al. 2018; Nelson et al. 2018), may then drive gas back out. Perhaps more importantly, these winds may set up a hot atmosphere in the circumgalactic regions that prevents further gas from coming in, bringing on quenching (e.g., Bower et al. 2017; Weinberger et al. 2017). Since massive galaxies around the Schechter mass reside in halos of $\log(M_h/M_\odot) \sim 12$ where virial shock-heating sets in (e.g., Kereš et al. 2005; Dekel & Birnboim 2006), the halo itself may become hot at the same time, further helping to slow down accretion of baryons onto the quenched galaxy, and thus possibly maintaining its quiescent state.

## 5. SUMMARY AND CONCLUSIONS

We have assembled a comprehensive set of nearly 600 near-IR spectra of $0.6<z<2.7$ galaxies covering the rest-optical H$\alpha$, [NII]$\lambda\lambda$6548,6584, [SII]$\lambda\lambda$6716,6731 line emission, and [OI]$\lambda$6300 for a fraction of them. The sample covers fairly homogeneously the bulk of the star-forming population over more than two orders of magnitude in stellar mass between $\log(M_*/M_\odot) \approx 9$ and 11.7, within roughly $\pm 0.6$ dex of the MS and $\pm 0.3$ dex of the mass-size relation of SFGs; it further extends well above and below the MS probing the regimes of "starburst outliers" and of quiescent populations, albeit with sparser sampling. Almost all (95%) of the sample is comprised of the sources observed in our KMOS$^{3D}$ (88%) and SINS/zC-SINF (7%) near-IR IFU surveys carried out at the ESO VLT (Wisnioski et al. 2015, 2018; Förster Schreiber et al. 2009, 2018; Mancini et al. 2011), and is benchmarked on the 3D-HST/CANDELS extragalactic surveys (Skelton et al. 2014; Momcheva et al. 2016; Koekemoer et al. 2011; Grogin et al. 2011). Galaxies taken from other (mostly long-slit) near-IR spectroscopic studies were added to boost the statistics at the high-mass end (Kriek et al. 2007; Newman et al. 2014; Wuyts et al. 2014; Barro et al. 2014; van Dokkum et al. 2015).

The sample size is nearly 6 times larger compared to our previous study of $z \sim 1 - 3$ galactic winds (G14), allowing us to establish trends in outflow incidence and physical properties more robustly. The KMOS and SINFONI IFU data sets are on average very deep (typical integrations of 8h on-source, and typical S/N ~ 15 per 30 km s$^{-1}$ channel in the integrated spectra), enabling us to investigate variations from stacked spectra sampling the parameter space more finely than before. Most importantly: (1) the primary stellar mass- and $K$-band magnitude selection of >90% of the sample minimizes biases towards subsets of the population with, e.g, higher-than-average AGN or star formation activity, thus providing a more complete census of ionized gas outflows and their characteristics; (2) the spatially-resolved IFU data available for almost all objects allow us to correct for the contribution from gravitational motions to the line profiles (traced via the narrow component velocity field) and therefore to optimally disentangle the emission from outflows down to faint/weak levels.

Motivated by the results from our previous work, we identified outflows from the detection of a broad (FWHM ≳ 400 km s$^{-1}$) emission component around the H$\alpha$+[NII] complex (and [SII] whenever possible), underneath a narrow star formation-dominated component, in the velocity-shifted galaxy spectra. Given the wide range in total line fluxes down to $\sim 10^{-17}$ erg s$^{-1}$ and the seeing-limited resolution (~0.5″, or ~ 4–5 kpc) of most of the data sets, we first classified outflows as being driven by an AGN if AGN activity is identified from X-ray/mid-IR/radio diagnostics and/or if the narrow [NII]$\lambda$6584/H$\alpha$>0.45. Different stacks of objects with thus-identified SF- or AGN-driven outflows exhibit the distinctive line widths and ratios between the two types seen in ~1–2 kpc resolution AO-assisted IFU data of several galaxies, in which the broad emission can be spatially associated with star-forming "clumps" in the disk or with the nuclear regions in AGN hosts (N12a; N12b; FS14). The resulting spectral characteristics in stacks indicates that this classification scheme is adequate, with little cross-contamination. About 1/4 of the galaxies drive an outflow, in roughly equal proportions between SF- and AGN-driven outflows (11% vs 15%, weighting secure and candidate outflows by 1 and 0.5).

Having identified the ionized gas outflows among the sample galaxies, we explored trends in incidence and physical properties, with the following main findings.

- In the log($M_*$)–$\Delta$MS plane, SF-driven and AGN-driven winds separate clearly. The incidence of SF-driven outflows varies primarily with $\Delta$MS and not with stellar mass. In contrast, the incidence of AGN-driven outflows only depends on stellar mass, and not on $\Delta$MS. More generally, the occurrence of SF-driven winds is closely coupled to star formation properties (specifically, $\Delta$MS, $\Sigma_{SFR}$, and the absolute and specific SFRs) whereas that of AGN-driven winds is strongly correlated with stellar mass and its central concentration ($\Sigma_*$, $n_{Sersic}$, and the inverse of $R_e$ where the structural parameters are fit to the rest-optical stellar light distribution). AGN-driven outflows are very rare at low masses but become common at high masses, across all $\Delta$MS, with fractions of 60%-75% at log($M_*/M_\odot$) $\gtrsim$ 10.8.

- For SF-driven winds, we find relatively low mass loading factors ($\eta \sim$ 0.1-0.2) for the ionized gas outflows, characterized by higher local electron densities than the HII regions in the galaxies. These low mass loading factors, and the rates of momentum and energy injection, could be easily accounted for by momentum-driven winds without invoking photon scattering. Although these properties imply that the warm ionized phase of the outflows would be insufficient to explain low galaxy formation efficiencies at low masses, they may represent only lower limits to the total outflow mass, momentum, and energy if it is mostly contained in a cold molecular and neutral phase and in a hot fluid as seen in $z \sim 0$ starbursts driving winds (e.g., Veilleux et al. 2005; Heckman & Thompson 2017). We also find that the SF-driven outflow velocities vary relatively little across the sampled mass range, resulting in a modest drop in $v_{out}/v_c$ consistent with the scenario in which winds may escape the galaxies at log($M_*/M_\odot$) $\lesssim$ 10.3 but only drive fountains at higher masses.

- The [NII]/H$\alpha$ excitation properties for the AGN-driven outflows along with the high inferred gas densities suggest that the wind emission comes from shocked cloudlets or filaments compressed and possibly entrained in a more tenuous hot wind fluid that is not probed by our observations. The rates of momentum and energy injection by these winds cannot be fully accounted for by the radiation from the star-forming populations and require an additional dominant driving mechanism, the AGN activity. Our estimates may again constitute lower limits as other wind phases are not included, or if the outflowing gas densities are overestimated. The high velocities of the AGN-driven winds allow escape from the galaxy. When these winds interact in the CGM and IGM with incoming gas from outside the halo, very high temperatures may be reached with long cooling time. This interaction could prevent the infalling gas from reaching the galaxy and thus contribute to shutting down star formation in the galaxies.

- The coincidence between the mass above which powerful AGN-driven winds become common and the Schechter mass, independently of redshift and star formation rate, seems too strong to be unrelated or caused by another hidden parameter or correlation. This conclusion is strengthened by the additional observational facts that galactic molecular gas fractions and specific SFRs drop with increasing log($M_*$) at roughly the same mass and similar rate as the AGN-driven outflows become more prominent. In a direct causal interpretation, this result would indicate that the central massive black hole in each galaxy accretes ever more rapidly as its mass increases with that of its host galaxy. The increased black hole accretion rates may lead to fast nuclear outflows ejecting mass from the nuclear regions as well as act as preventive feedback in the circum-/intergalactic medium.

Given the multi-phase nature of outflows, and the obvious importance of determining reliable mass outflow rates and mass loading factors, it is highly desirable to improve on the quantitative constraints of observations and predictions of simulations. A key challenge from the observational side remains the determination of the wind geometry and the physical conditions of the outflowing gas, both critical in estimating $\dot{M}_{out}$ and $\eta$. Very sensitive, high spatial and spectral resolution mapping of ionized and even molecular gas winds at $z \sim 1 - 3$ exist but for a limited number of normal galaxies (e.g., N12a; FS14; Herrera-Camus et al. 2018) and luminous but rare QSOs (e.g., Cresci et al. 2015; Perna et al. 2015; Carniani et al. 2015, 2016; Brusa et al. 2016, 2018; Kakkad et al. 2016). Future progress will benefit greatly from further such studies, covering better typical galaxies and ideally combining multi-phase outflow probes.

We thank the ESO Paranal staff for their excellent support throughout the many observing runs during which the KMOS and SINFONI data used in this work were taken. We also thank V. Arumugam, R. J. Ivison, and D. J. Rosario for sharing information in advance of publication used in our AGN identification, as well as N. Herrera-Ruiz for providing VLBA information to that purpose. We are grateful to our colleagues for many stimulating and helpful discussions on various aspects of this paper, especially C. Steidel, M. Salvato, and A. E. Shapley. We thank the referee for constructive and useful suggestions for improvement. E. S. W. and J. T. M. acknowledge support by the Australian Research Council Center of Excellence for All Sky Astrophysics in 3 Dimensions (ASTRO 3D), through project number CE170100013. D. J. W. and M. F. acknowledge the support of the Deutsche Forschungsgemeinschaft via Project ID 3871/1-1 and 3871/1-2. G. B. B. acknowledges support from the Cosmic Dawn Center, which is funded by the Danish National Research Foundation. The LBT is an international collaboration among institutions in the United States, Italy, and Germany. LBT Corporation partners are: The University of Arizona on behalf of the Arizona university system; Istituto Nazionale di Astrofisica, Italy; LBT Beteiligungsgesellschaft, Germany, representing the Max-Planck Society, the Astrophysical Institute Potsdam, and Heidelberg University; The Ohio State University, and the Research Corporation, on behalf of The Universities of Notre Dame, Minnesota, and Virginia.

**Table 1.** Derived Physical Properties of SF- and AGN-Driven Outflows

| Stack | $\langle z \rangle$ | $\langle \log(M_*/M_\odot) \rangle$ | $\langle \Delta MS \rangle$ | $L(H\alpha)_{0,SF}$ (erg s$^{-1}$) | $\langle SFR \rangle$ (M$_\odot$ yr$^{-1}$) | $\langle R_e \rangle$ (kpc) | $F_{br}/F_{na}(H\alpha)$ | $L(H\alpha)_{0,br}$ (erg s$^{-1}$) | $v_{out}$ (km s$^{-1}$) | $\langle v_c \rangle$ (km s$^{-1}$) | $v_{out}/v_c$ | $R_e/v_{out}$ (yr) | $n_{e,br}$ (cm$^{-3}$) | $M_{out}$(HII+He) (M$_\odot$) | $\dot{M}_{out}$ (M$_\odot$ yr$^{-1}$) | $\eta$ | $P_{out}/P_{rad}$ | $\dot{E}_{out}/L_{SFR}$ |
|---|---|---|---|---|---|---|---|---|---|---|---|---|---|---|---|---|---|---|
| (1) | (2) | (3) | (4) | (5) | (6) | (7) | (8) | (9) | (10) | (11) | (12) | (13) | (14) | (15) | (16) | (17) | (18) | (19) |
| *Star Formation-Driven Outflows* | | | | | | | | | | | | | | | | | | |
| Best spectra (33, secure detections) | 1.84 | 10.58 | 0.180 | 9.9E+42 | 47.3 | 3.4 | 0.68 | 6.75E+42 | 450 | 220 | 2.0 | 7.7E+06 | $380^{+249}_{-167}$ | 5.77E+07 | 7.5 | 0.16 | 1.0 | 7.8E-04 |
| z<1.7 (35, incl. candidates) | 0.92 | 10.21 | 0.350 | 6.2E+42 | 29.3 | 3.1 | 0.63 | 3.88E+42 | 460 | 200 | 2.3 | 6.7E+06 | 380 | 3.31E+07 | 4.9 | 0.17 | 1.1 | 8.7E-04 |
| z>1.7 (49, incl. candidates) | 2.27 | 10.55 | 0.160 | 2.0E+43 | 94.0 | 3.2 | 0.88 | 1.74E+43 | 460 | 190 | 2.4 | 6.9E+06 | 380 | 1.48E+08 | 21.5 | 0.23 | 1.5 | 1.2E-03 |
| log($M_*/M_\odot$)<10 (18, incl. candidates) | 1.00 | 9.71 | 0.410 | 3.2E+41 | 1.5 | 2.5 | 0.54 | 1.70E+41 | 260 | 119 | 2.2 | 9.6E+06 | 380 | 1.45E+06 | 0.2 | 0.10 | 0.4 | 1.7E-04 |
| log($M_*/M_\odot$)=10-10.3 (12, incl. candidates) | 2.20 | 10.17 | 0.180 | 8.2E+42 | 39.0 | 2.9 | 0.76 | 6.22E+42 | 490 | 139 | 3.5 | 6.0E+06 | 380 | 5.32E+07 | 8.9 | 0.23 | 0.6 | 4.5E-04 |
| log($M_*/M_\odot$)=10.3-10.6 (18, incl. candidates) | 2.20 | 10.49 | 0.100 | 1.1E+43 | 51.0 | 3.1 | 0.74 | 7.93E+42 | 400 | 241 | 1.7 | 7.7E+06 | 380 | 6.77E+07 | 8.8 | 0.17 | 0.3 | 2.3E-04 |
| log($M_*/M_\odot$)=10.6-10.9 (19, incl. candidates) | 2.00 | 10.74 | 0.210 | 1.8E+43 | 88.0 | 4.1 | 0.83 | 1.53E+43 | 410 | 264 | 1.6 | 1.0E+07 | 380 | 1.31E+08 | 13.0 | 0.15 | 0.7 | 5.1E-04 |
| log($M_*/M_\odot$)>10.9 (11, incl. candidates) | 2.20 | 11.05 | 0.270 | 5.3E+43 | 253.0 | 4.4 | 0.39 | 2.07E+43 | 430 | 304 | 1.4 | 1.0E+07 | 380 | 1.77E+08 | 17.1 | 0.07 | 0.3 | 2.1E-04 |
| *AGN-Driven Outflows* | | | | | | | | | | | | | | | | | | |
| Best spectra (30, secure detections) | 1.60 | 11.00 | 0.160 | 2.0E+43 | 97.0 | 2.9 | 0.86 | 1.75E+43 | 2300 | 284 | 8.1 | 1.3E+06 | 1000 | 5.69E+07 | 44.7 | 0.46 | 15.4 | 5.9E-02 |
| z<1.7 (62, incl. candidates) | 1.31 | 10.95 | 0.150 | 8.8E+42 | 42.0 | 3.8 | 0.68 | 6.00E+42 | 1990 | 283 | 7.0 | 1.9E+06 | $915^{+605}_{-339}$ | 1.95E+07 | 10.2 | 0.24 | 7.0 | 2.3E-02 |
| z>1.7 (36, incl. candidates) | 2.32 | 11.05 | -0.050 | 3.1E+43 | 148.0 | 2.9 | 0.68 | 2.11E+43 | 1000 | 286 | 3.5 | 2.9E+06 | $1229^{+579}_{-357}$ | 6.86E+07 | 23.4 | 0.16 | 2.3 | 3.8E-03 |
| [NII]/H$\alpha_{na}$=0.45-0.75 (54, incl. candidates) | 1.61 | 11.00 | 0.010 | 1.7E+43 | 83.0 | 3.3 | 0.31 | 5.40E+42 | 1400 | 309 | 4.5 | 2.4E+06 | $659^{+1811}_{-474}$ | 1.75E+07 | 7.3 | 0.09 | 1.8 | 4.2E-03 |
| [NII]/H$\alpha_{na}$>0.75 (35, incl. candidates) | 1.03 | 11.03 | -0.110 | 9.0E+42 | 43.0 | 3.1 | 0.52 | 4.70E+42 | 900 | 274 | 3.3 | 3.5E+06 | >3375 | 1.53E+07 | 4.4 | 0.10 | 1.3 | 2.0E-03 |
| "Strong Outflows" (8) | 1.60 | 10.97 | 0.630 | 5.3E+43 | 253.0 | 3.1 | 1.34 | 7.11E+43 | 1500 | 301 | 5.0 | 2.1E+06 | 1000 | 2.31E+08 | 111.7 | 0.44 | 9.6 | 2.4E-02 |

NOTE — Properties are reported for subsets of galaxies and their co-averaged spectra, as listed in Column 1, where the number of galaxies included is given in parenthesis. Galaxy properties include the following, corresponding to the median value for the galaxies entering each stacked spectrum: redshift (Column 2), stellar mass log($M_*$), MS offset $\Delta MS$, SFR, and effective radius $R_e$ (Columns 3, 4, 6, and 7), the total extinction-corrected H$\alpha$ luminosity (Column 5), and the circular velocity derived from the measured kinematics (Column 11). Outflow properties include the broad-to-narrow H$\alpha$ flux ratio and implied intrinsic H$\alpha$ luminosity of the outflowing ionized gas (Columns 8 and 9), the outflow velocity from the broad component line width (Column 10), the ratio of outflow to circular velocities (Column 12), the time for the outflow to travel a distance of $R_e$ at constant velocity (Column 13), the local electron densities derived from the [SII]$\lambda$6716/$\lambda$6731 flux ratio in the broad component (Column 14), the mass and mass ejection rate of the outflowing ionized gas including H and He (Columns 15 and 16), the mass loading $\eta = \dot{M}_{out}$/SFR, the outflow to stellar radiation momentum ratio, and the outflow kinetic energy to stellar luminosity ratio (Columns 17, 18, and 19). For $n_{e,br}$, the fitting results are listed if the [SII] ratio in the broad component was measured at >3$\sigma$, with uncertainties corresponding to the 68% confidence intervals if the ratio is >1$\sigma$ away from the high-density limit (adopting the calculations by Sanders et al. 2016). If the [SII] ratio was confidently measured but within 1$\sigma$ of the high-density regime, the lower 68% confidence interval is given as limit. If the fitted FWHM of the broad component exceeded 1500 km s$^{-1}$, roughly twice the [SII] line separation, the fits become questionable and we then list the best-determined estimate from the successful fits for either SF-driven or AGN-driven outflow stacks (see text). For all calculations of mass outflow, momentum, and energy rates, we used a common $n_{e,br}$ = 380 cm$^{-3}$ for the SF-driven outflows, and 1000 cm$^{-3}$ for the AGN-driven outflows (see Sections 4.1 and 4.2).

# APPENDIX

## A. REFERENCES FOR AGN IDENTIFICATION FROM ANCILLARY DATA

This Section lists the source catalogs and references used for the identification of AGN among the KMOS$^{3D}$, LUCI, and vD15/B14 subset of the full sample. The identification was based on X-ray, mid-IR, and radio indicators, wide-field VLBI techniques, and source variability searches. Source matching to the optical positions of the objects in our sample was done using a typical radius of 0.5″ to 1″, except for the radio continuum surveys where a typical radius of 2″ was used. All luminosities required (hard 2-10 keV X-ray or radio) were computed from the catalog fluxes adopting the H$\alpha$ redshift of the galaxies. For SINS/zC-SINF and K07 objects, we adopted the identifications reported by FS09, Mancini et al. (2011), Förster Schreiber et al. (2018), and Kriek et al. (2007,2008), to which we refer for details.

### A.1. GOODS-South Field

• X-ray: 7Ms Chandra Deep Field South survey catalog of Luo et al. (2017).
• Mid-IR: Spitzer/IRAC photometry from the CANDELS GOODS-S multi-wavelength catalog of Guo et al. (2013).
• Radio: VLA 1.4 GHz catalog of Miller et al. (2013).
• VLBI: VLBA 1.4 GHz survey of Middleberg et al. (2011).
• Variability: Optical searches with the ESO-MPI 2.2m telescope by Trevese et al. (2008), with HST by Villforth et al. (2010) and Sarajedini et al. (2011), and with the VLT Survey Telescope (VST) by Falocco et al. (2015). X-ray search with Chandra by Young et al. (2012).

### A.2. COSMOS Field

• X-ray: Chandra 4.6 Ms COSMOS Legacy Survey of Civano et al. (2016).
• Mid-IR: Spitzer/IRAC photometry from the COSMOS2015 catalog by Laigle et al. (2016).
• Radio: VLA-COSMOS 3.0 GHz Large Project catalog of Smolčić et al. (2017).
• VLBI: VLBA 1.4 GHz survey of Herrera-Ruiz et al. (2017; and priv. comm.).
• Variability: Optical search with the VST by de Cicco et al. (2015).

### A.3. UDS Field

• X-ray: Chandra X-UDS Survey catalog of Kocevski et al. (2018; data kindly provided by D. J. Rosario, priv. comm.).
• Mid-IR: Spitzer/IRAC photometry from the CANDELS UDS multi-wavelength catalog of Galametz et al. (2013).
• Radio: VLA 1.4 GHz 7$\mu$Jy catalog of V. Arumugam et al. (in preparation; data kindly provided by R. J. Ivison, priv. comm.).

### A.4. GOODS-North Field

• X-ray: 2Ms Chandra Deep Field North Survey catalog of Xue et al. (2016).
• Radio: VLA 1.4 GHz catalog of Morrison et al. (2010).

### A.5. EGS Field

• X-ray: Chandra AEGIS-X Survey catalog of Nandra et al. (2015).
• Mid-IR: Spitzer/IRAC photometry from the CANDELS EGS multi-wavelength catalog of Stefanon et al. (2017).
• Radio: VLA 1.4 GHz AEGIS20 Survey catalog of Ivison et al. (2007).